\documentclass[twocolumn]{aastex631}

\usepackage{graphicx}
\usepackage{amsmath}	
\usepackage{amssymb}
\usepackage{booktabs}
\usepackage{enumerate}
\usepackage{dirtytalk}
\usepackage{float}
\usepackage{enumitem}
\usepackage{calligra}
\usepackage{booktabs}
\usepackage{lipsum} 
\usepackage{float}
\usepackage{xcolor}
\usepackage{multirow}
\usepackage{makecell}
\usepackage{blindtext}
\usepackage{subfigure}
\usepackage{gensymb}

\shorttitle{ROXs 42B b}
\shortauthors{Inglis et al.}

\begin{document}

\title{Atmospheric Retrievals of the Young Giant Planet ROXs 42B b from Low- and High-Resolution Spectroscopy}

\correspondingauthor{Julie Inglis}
\email{jinglis@caltech.edu}

\author[0000-0001-9164-7966]{Julie Inglis}
\affiliation{Division of Geological and Planetary Sciences, California Institute of Technology, Pasadena, CA, 91125}

\author[0000-0003-0354-0187]{Nicole L. Wallack}
\affiliation{Earth and Planets Laboratory, Carnegie Institution for Science, Washington, DC, 20015}

\author[0000-0002-6618-1137]{Jerry W. Xuan}
\affiliation{Department of Astronomy, California Institute of Technology, Pasadena, CA, 91125}

\author[0000-0002-5375-4725]{Heather A. Knutson}
\affiliation{Division of Geological and Planetary Sciences, California Institute of Technology, Pasadena, CA, 91125}

\author[0000-0003-1728-8269]{Yayaati Chachan}
\altaffiliation{CITA National Fellow}
\affiliation{Department of Physics and Trottier Space Institute, McGill University, 3600 rue University, H3A 2T8 Montreal QC, Canada}
\affiliation{Trottier Institute for Research on Exoplanets (iREx), Université de Montréal, Canada}

\author[0000-0002-6076-5967]{Marta L. Bryan}
\affiliation{Department of Astronomy and Astrophysics, University of Toronto, Toronto, Ontario, Canada M5S 3H4}

\author[0000-0003-2649-2288]{Brendan P. Bowler}
\affiliation{Department of Astronomy, The University of Texas at Austin, Austin, TX, 78712}

\author[0000-0003-0971-1709]{Aishwarya Iyer}
\altaffiliation{NASA Postdoctoral Fellow}
\affiliation{Goddard Space Flight Center, Greenbelt, MD 20771}

\author[0000-0003-3759-9080]{Tiffany Kataria}
\affiliation{Jet Propulsion Laboratory, California Institute of Technology, Pasadena, CA 91109}

\author[0000-0001-5578-1498]{Bj\"{o}rn Benneke}
\affiliation{Department of Physics and Trottier Institute for Research on Exoplanets, Université de Montréal, Montreal, QC, Canada}

\begin{abstract}

Previous attempts have been made to characterize the atmospheres of directly imaged planets at low-resolution (R$\sim$10s-100s), but the presence of clouds has often led to degeneracies in the retrieved atmospheric abundances with cloud opacity and temperature structure that bias retrieved compositions. In this study, we perform retrievals on the ultra-young ($\lesssim$ 5 Myr) directly imaged planet ROXs 42B b with both a downsampled low-resolution $JHK$-band spectrum from Gemini/NIFS and Keck/OSIRIS, and a high-resolution $K$-band spectrum from pre-upgrade Keck/NIRSPAO. Using the atmospheric retrieval framework of \texttt{petitRADTRANS}, we analyze both data sets individually and combined. We additionally fit for the stellar abundances and other physical properties of the host stars, a young M spectral type binary, using the SPHINX model grid. We find that the measured C/O, $0.50\pm0.05$, and metallicity, [Fe/H] = $-0.67\pm0.35$, for ROXs 42B b from our high-resolution spectrum agree with that of its host stars within 1$\sigma$. The retrieved parameters from the high-resolution spectrum are also independent of our choice of cloud model. In contrast, the retrieved parameters from the low-resolution spectrum show strong degeneracies between the clouds and the retrieved metallicity and temperature structure. When we retrieve on both data sets together, we find that these degeneracies are reduced but not eliminated, and the final results remain highly sensitive to cloud modeling choices. We conclude that high-resolution spectroscopy offers the most promising path for reliably determining atmospheric compositions of directly imaged companions independent of their cloud properties.
\end{abstract}

\section{Introduction} \label{sec:intro}

\begin{deluxetable}{cll}
\tablecaption{Summary of the previously derived system parameters and data used in our analysis of ROXs 42B b along with new derived parameters for the host star from this work using the SPHINX model grid from \cite{iyer_sphinx_2023}. \label{table:systemvals}}
\tabletypesize{\footnotesize}
\tablehead{\colhead{Parameter} & \colhead{Value} & \colhead{Reference}}
     \startdata
     \hline
     \multicolumn{3}{c}{Primary (ROXs 42B AB)}  \\
     \hline
     RA (J2000) & 16 31 15.018 & \cite{gaia_collaboration_vizier_2022}\\
     Dec (J2000) & -24 32 43.715 & \cite{gaia_collaboration_vizier_2022} \\
     SpT & M0$\pm1$ & \cite{bowler_2014} \\
     $m_J$ (mag) & $9.91\pm0.02$ & \cite{bowler_2014} \\
     $m_H$ (mag) & $9.02\pm0.02$ & \cite{bowler_2014}\\
     $m_{K_s}$ (mag) & $8.67\pm0.02$ & \cite{bowler_2014} \\
     $m_{L'}$ (mag) & $8.42\pm0.05$ & \cite{daemgen_2017} \\
     $A_V$ & $1.7^{+0.9}_{-1.2}$ & \cite{bowler_2014} \\
     dist (pc) & $146.447_{-0.6627}^{+0.6627}$ & \cite{gaia_collaboration_vizier_2022}\\
     Age (Myr) & $6.8^{+3.4}_{-2.3}$ & \cite{kraus_2013} \\
     $M_1$ (M$_\odot$) & $0.89\pm0.08$ & \cite{kraus_2013}\\
     $M_2$ (M$_\odot$) & $0.36\pm0.04$ & \cite{kraus_2013}\\
     $T_{eff}$  (K) & $3850\pm80^a$ & \cite{kraus_2013}\\
     binary sep (mas) & $\approx83$ & \cite{kraus_2013}\\
     flux ratio ($\Delta K$) & $\approx1.1$ & \cite{kraus_2013}\\ 
     $R_1$ (R$_\odot$) & $1.51\pm0.02$ & This work  \\
     $R_2$ (R$_\odot$) & $1.39\pm0.02$ & This work  \\
     $T_{eff,1}$  (K) & $3650\pm20$ & This work  \\
     $T_{eff,2}$  (K) & $2600\pm20$ & This work  \\
     log \textit{g}$_1$ (cgs) & $4.12\pm0.1$ & This work  \\
     log \textit{g}$_2$ (cgs) & $4.26\pm0.1$ & This work  \\
     $M_1$ ([M$_{\odot}$) & $1.12\pm0.13$ & This work  \\
     $M_2$ (M$_{\odot}$) & $1.29\pm0.26$ & This work \\
     C/O & $0.54\pm0.01$ & This work  \\
     $[\text{Fe/H}]$ & $-0.30\pm0.03$ & This work  \\
     $A_V$ & $2.1\pm0.1$ & This work \\    
     \hline
     \multicolumn{3}{c}{Companion (ROXs 42B b)}  \\ 
     \hline
     SpT & L1$\pm$1 & \cite{bowler_2014} \\
     $m_J$ (mag) & $16.91\pm0.11$ & \cite{kraus_2013} \\
     $m_H$ (mag) & $15.88\pm0.05$ & \cite{kraus_2013} \\
     $m_{K_s}$ (mag) & $15.01\pm0.06$ & \cite{kraus_2013}\\ 
     $m_{L'}$ (mag) & $13.97\pm0.06$ & \cite{daemgen_2017}\\ 
     $m_{Br\alpha}$ (mag) & $13.90\pm0.08$ & \cite{daemgen_2017}\\ 
     $m_{M_s}$ (mag) & $14.01\pm0.23$ & \cite{daemgen_2017}\\ 
     sep ($"$) & 1.170 & \cite{kraus_2013} \\
     sep (au) & 175 & \cite{kraus_2013}$^b$\\
     $M$ (M$_{J}$) & $10\pm4$ & \cite{kraus_2013} \\
     Age (Myr) & $3\pm2$ & \cite{bryan_2018} \\
     \textit{v sini} & $9.5^{+2.1}_{-2.3}$ & \cite{bryan_2018} \\
     & 1800–2600 K & \cite{currie_direct_2013} \\
     & (K-band) & \\
     $T_{eff}$ & 1950-2000 & \cite{currie_2014} \\
     & (photometry) & \\
     & 1600-2000 K & \cite{daemgen_2017} \\
     & (photometry) & \\ \hline
    \enddata
    \tablecomments{a) The reported $T_{eff}$~corresponds to the primary, brighter component in the binary, calculated using the reported $K$-band flux ratio in \cite{ratzka_2005}. b) Calculated using the separation measured by \cite{kraus_2013} and the updated distance from \cite{gaia_collaboration_vizier_2022}}
\end{deluxetable}

Directly imaged planets are a population of massive, self-luminous companions, typically observed from orbital separations comparable to that of Saturn, to well beyond the orbital distance of Neptune (10s-100s of au). These companions are typically young, with ages $\lesssim$ 100 Myr, and are often distinguished from brown dwarfs companions  ($\sim13-80$ M$_{J}$) as having masses near or below the deuterium-burning limit at $\sim$13 M$_{J}$ \citep{bowler_imaging_2016,currie_direct_2022}. A more physically motivated distinction between directly imaged planets and brown dwarfs would be based on formation process, as planets are expected to form bottom up from core accretion, while brown dwarf companions form from gravitational collapse, however this is challenging to distinguish for individual objects \citep{pollack_formation_1996,boss_giant_1997,alibert_models_2005}.

Population studies indicate that these lower mass companions have properties that are distinct from those of massive brown dwarf companions. Planetary-mass companions appear to have a higher occurrence rate than brown dwarf companions to the same stellar population, and tend to have lower orbital eccentricities \citep{nielsen_gemini_2019,wagner_mass_2019,bowler_population-level_2020,nagpal_impact_2023}, and distinct stellar spin-orbit orientations \citep{bowler_rotation_2023}. The differing properties of these two populations suggest that they likely have distinct formation channels, but the nature of these formation channels is currently debated. It has also been suggested that one or both populations might not have formed in situ, but instead migrated away from their initial formation locations \citep{scharf_long-period_2009}. However, \cite{bryan_searching_2016} failed to find evidence for additional massive bodies in systems hosting wide-separation directly imaged companions that could have acted as scatterers, suggesting this is not the dominant formation mechanism for these planets.

Numerous studies have explored how the elemental abundances of giant planet atmospheres, which are often parameterized as a carbon-to-oxygen ratio (C/O) and bulk metallicity (e.g. [Fe/H]), can encode information about their formation and migration histories \citep[e.g.,][]{madhusudhan_co_2012,oberg_effects_2011,cridland_composition_2016,mordasini_imprint_2016,turrini_tracing_2021}. 
 Companions that form via gravitational instability, including cloud fragmentation and disk instability, are expected to have envelope compositions that are largely similar to those of their host stars, (analogous to binary star formation), although is has been shown that disk structures can result in local enhancements of solid material that can be incorporated during collapse \citep{boley_possibility_2010,boley_heavy-element_2011}. Alternatively, core accretion \citep{pollack_formation_1996,lambrechts_rapid_2012} is expected to produce giant planets with a wide range of C/O and bulk metallicity values. This is because the compositions of the solid and gaseous components of the disk vary as a function of time and location within the disk, and different planets are expected to accrete different quantities of solids versus gas \citep{oberg_effects_2011,turrini_tracing_2021,chachan_breaking_2023}. However, since collapse from instability is more likely to occur early in the disk's lifetime \citep{boley_two_2009}, a newly formed planet could still have access to a substantial reservoir of material in the disk. If the planet is able to accrete additional material following initial collapse phase, a range of envelope compositions may be possible for planets formed via disk instability as well. Nonetheless, atmospheric composition could still probe the formation and migration histories of directly imaged planets and help distinguish between different formation scenarios.
 
Atmospheric characterization analyses of directly imaged planets have historically made use of photometric and low-resolution spectroscopic data sets (resolving power $R\sim20-100$), which are compared to grids of self-consistent, 1-dimensional radiative-convective equilibrium forward models \citep[e.g. the early studies of the HR 8799 planets:][]{galicher_m-band_2011,barman_young_2011,currie_combined_2011,marley_masses_2012}. These models have relatively limited degrees of freedom, and often produce poor fits to these spectra.  More recently, a number of studies have begun to utilize spectral inversion, or atmospheric retrieval, techniques to model the atmospheres of isolated brown dwarfs \citep{line_2017,burningham_2017,burningham_2021,lueber_2022,calamari_atmospheric_2022,hood_brown_2023}, directly imaged companions \citep{molliere_2020,ruffio_deep_2021,wang_retrieving_2022,zhang_elemental_2023}, and brown dwarf companions \citep{xuan_clear_2022}. These atmospheric retrievals can be used to place constraints on the C/O ratios and metallicities of these objects, as well as other physical quantities such as their radii, surface gravities, and cloud properties. 

\begin{figure*}[t!]
\plotone{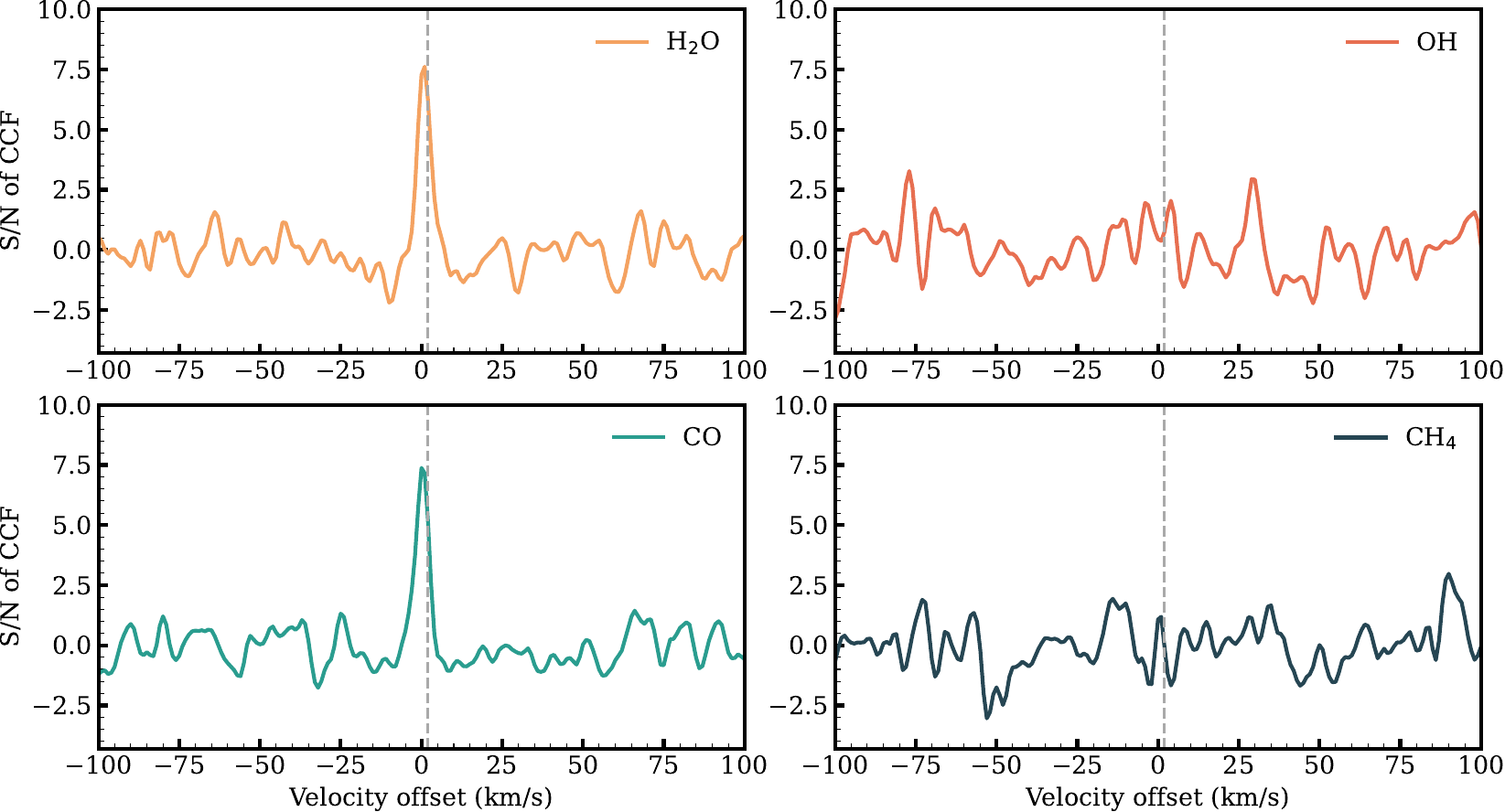}
\caption{Cross-correlation functions (CCFs) showing detections of CO and H$_2$O using two orders ($2.27-2.38$ $\mu$m) from the high-resolution NIRSPEC $K$-band spectrum. The significance is computed by dividing the CCF by the standard deviation of the out-of-peak baseline.
\label{fig:ccf}}
\end{figure*}

Many of the directly imaged planets and brown dwarfs examined in the literature lie in a temperature regime where silicates are expected to condense in their upper atmospheres \citep{marley_clouds_2015}.  The presence of these clouds has a strong effect on the observed shapes of their low-resolution emission spectra \citep[e.g.,][]{burningham_2017}. Previous studies have explored a wide range of cloud models with varying degrees of physical motivation and success at reproducing the observed features. They also revealed that low-resolution retrievals often exhibit strong degeneracies between their retrieved cloud properties and other physical parameters, such as effective temperature and metallicity. For example, a retrieval analysis by \cite{burningham_2017} found that their cloud-free models tended towards isothermal pressure-temperature (P-T) profiles and high metallicities in order to compensate for the effects of clouds. This degeneracy has been noted in many subsequent studies, with a variety of proposed solutions \cite[e.g.][]{piette_considerations_2020,zhang_elemental_2023}. \cite{lueber_2022} found that their treatment of cloud opacity also resulted in trade offs with the retrieved surface gravity, and were unable to obtain good constraints on the C/O ratios of many of the brown dwarfs in their sample. On the other hand, \cite{molliere_2020} retrieved a C/O value consistent with solar in their paper on HR 8799 e. This value was independent of the cloud model used and did not correlate with other physical parameters such as the vertical temperature structure.

We can obtain a complementary perspective on the atmospheres of these cloudy objects using high-resolution spectroscopy. At high spectral resolution, the ratios of the line depths of different molecular species can be used to constrain their relative abundances in a way that is less biased by the presence of continuum opacity sources from clouds. Since the cores of absorption lines are emitted higher up in the atmosphere than the continuum flux, high-resolution spectroscopy also allows observers to probe a much broader range of pressures, including pressures that lie above the cloud tops. This can increase the signal-to-noise of molecular detections for cloudy objects \citep{gandhi_seeing_2020, hood_prospects_2020}.  In their recent study of the brown dwarf companion HD 4747 B, \cite{xuan_clear_2022} found their retrieved abundances from high-resolution spectroscopy were insensitive to their choice of cloud model. Similarly, \cite{wang_retrieving_2022} performed joint retrievals on low and high-resolution spectroscopy of HR~8799~e and found that while clouds are likely present in this planet's atmosphere, they are located deeper in the atmosphere than where the majority of the flux was emitted, resulting in a minimal effect on their observed spectrum. They obtained good constraints on the C/O, C/H and O/H ratios, retrieving values that were consistent with other studies of the HR 8799 planets \citep[e.g.][]{ruffio_deep_2021} and AF Lep b \citep{zhang_elemental_2023}.

ROXs 42B b is an ultra-young directly imaged planetary companion on a wide orbit around a low mass binary star. Its large separation ($\sim$1") from its host stars makes it a favorable target for atmospheric characterization. It's large orbital separation and low-mass binary host make it an interesting case study for formation. In this paper, we perform atmospheric retrievals on low- and high-resolution spectroscopy of the young companion ROXs 42B b.  \cite{bowler_2014} published a medium-resolution spectrum for this object spanning $J$-, $H$- and $K$-bands (1.1 - 2.4 $\mu$m), while \cite{bryan_2018} published a high-resolution $K$-band spectrum.  However, neither of these studies sought to interpret these data using atmospheric retrieval modeling. We use the open-source radiative transfer code \texttt{petitRADTRANS} \citep{molliere_petitradtrans_2019, molliere_2020} in order to constrain the properties of ROXs 42B b's atmosphere, including its C/O ratio and metallicity. We compare our constraints from low-resolution only (\S\ref{sec:Low_res}), high-resolution only (\S\ref{sec:High_res}), and joint high- and low-resolution retrievals (\S\ref{sec:Joint_Ret}) to explore the advantages and disadvantages of each data type in the context of atmospheric characterization. Finally, we compare our retrieved abundances for ROXs 42B b to those of its primary, which is an unresolved binary, in \S\ref{sec:hoststar} and discuss implications for the formation of this system.

\section{System Properties}

ROXs 42B is a close binary system\footnote{ROXs 42A and ROXs 42C are two nearby stars located within the error circle of the Einstein Observatory which first identified the Rho Ophiuchus X-ray source, ROXs 42 \citep{montmerle_1983}. While ROXs 42C has a similar age to ROXs 42B, it has a very different mass and is widely separated from ROXs 42B, so is unlikely to be affiliated, while ROXs 42A is likely a field star \citep{bouvier_1992}.} in the $\rho$ Ophiuchus star-forming region \citep{bouvier_1992}. A candidate companion was first identified around the binary by \cite{ratzka_2005}. The companion was later confirmed by \cite{kraus_2013} and \cite{currie_direct_2013}. \cite{kraus_2013} measured a projected separation of 1.17", which corresponds to a separation of 175 au with the updated distance from \cite{gaia_collaboration_vizier_2022}. The primary binary, ROXs 42B AB, has a (unresolved) spectral type of M0 and has been identified as a young system based on weak H$\alpha$ emission, X-ray emission, and possible lithium absorption in its optical spectrum \citep{bouvier_1992}. Isochronal fitting gives an approximate stellar age of 6.8$^{+3.4}_{-2.3}$ Myr \citep{kraus_2013}. Models suggest individual component masses of $0.89\pm0.08$ and $0.36\pm 0.04$ $M_{\odot}$ \citep{kraus_2013}. The absence of infrared excess in either $W$3 or $W$4 in WISE observations, or a radio emission at 1.3 mm, indicates that this system does not contain a significant protoplanetary disk \citep{kraus_2013,daemgen_2017}. However, Spitzer observations at 8 $\mu$m suggest that the companion, ROXs 42B b has excess infrared emission, and might therefore host a circumplanetary disk despite the lack of accretion signatures \citep{martinez_mid-infrared_2021}. ROXs 42B b is closest to a spectral type of L0, but is also observed to have colors much redder than field objects of a similar spectral type, which could be due to circumstellar dust \citep{kraus_2013,daemgen_2017}.

\begin{figure*}[th!]
\epsscale{1.1}
\plotone{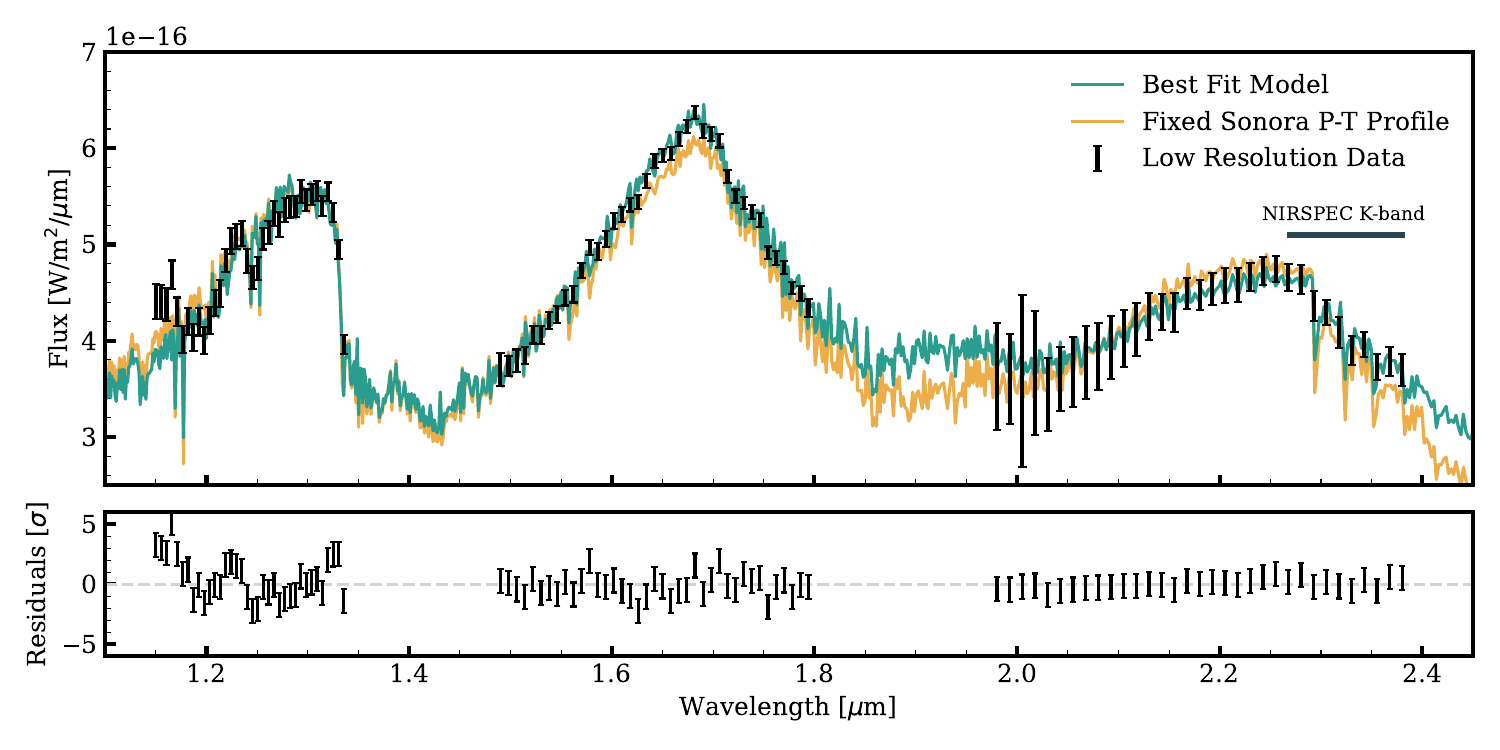}
\caption{Retrieved best-fit model (green) compared to our low-resolution $JHK$-band spectrum from \cite{bowler_2014} (top panel) and error-weighted residuals (lower panel). Our full low-resolution retrievals are compared to the best fit model (yellow) from our retrievals where our P-T profile is fixed to a Sonora Bobcat P-T profile from \citep{marley_sonora_2021}. The horizontal black line indicates the wavelength coverage of our Keck/NIRSPEC high-resolution $K$-band data for comparison.
\label{fig:LR_spec}}
\end{figure*} 

Evolutionary model fits to photometry of ROXs 42B b place approximate bounds on its mass of $10\pm4$ M$_{J}$ for an assumed age between $1-5$ Myr \citep{kraus_2013} using models from \cite{chabrier_evolutionary_2000}. In a follow-on study, \cite{bowler_2014} obtained medium-resolution near infrared (NIR) $JHK$-band spectra of the companion and found that the shape of the $H$-band flux peak was best-matched by grid models with low surface gravities, consistent with its relatively young age. Published constraints on the effective temperature of this object vary. Values from the photometric study of \cite{currie_2014} ranged from $1800-2200$ K depending on the models used and whether or not clouds were included. \cite{currie_direct_2013} fit a low-resolution $K$-band spectrum from VLT/SINFONI and found a significantly higher effective temperature range of 1800-2600 K, which also varied depending on their treatment of clouds in the grid models.  There are currently no published constraints on ROXs 42B b's atmospheric metallicity or C/O ratio, due to the limitations of fits to publicly available grid models  \citep{currie_2014,daemgen_2017}. \cite{bryan_2018} obtained a high-resolution $K$-band spectrum for this object and used it to measure a projected rotational velocity of 9.5$^{+2.1}_{-2.3}$ km s$^{-1}$, but this study assumed a solar composition atmosphere. To date, the only published composition constraint for this object is an upper limit on its CO$_2$ abundance from \cite{daemgen_2017}.

There are currently no published composition constraints for the primary, ROXs 42B AB, either. Only a few directly imaged planet hosts in the literature have measured abundances, e.g. HR 8799 \citep{wang_chemical_2020}. In most cases, the retrieved chemical abundances are compared to solar values instead of their stellar value. However, in order to make determinations about the formation mechanisms of the companion, it is necessary to have stellar properties to compare to our retrieved abundances for the companion. In Section~\ref{sec:hoststar}, we fit the optical SNIFS spectrum of the (unresolved) primary ROXs 42B AB from \cite{bowler_2014} using the SPHINX M dwarf model grid from \cite{iyer_sphinx_2023} to constrain the abundances of the host binary.

\section{Archival Data Used in This Study} \label{sec:Data}

\subsection{Low-resolution Spectroscopy of ROXs 42B b}

\cite{bowler_2014} published a medium resolution (R$\sim$1000) spectrum of ROXs 42B b. The $H$-band data in this study were taken on UT May 9, 2012 and the $J$-band data were taken on UT May 14 and July 4, 2012 using ALTAIR, the facility AO system, and the Near-Infrared Integral Field Spectrometer \citep[NIFS,][]{mcgregor_gemini_2003}, an image-slicer integral field spectrograph with a resolution of R$\sim$5000. The $K$-band data were taken with the OH-Suppressing Infrared Imaging Spectrograph \citep[OSIRIS,][]{larkin_osiris_2006} on Keck II on UT August 20 2011. The OSIRIS measurements used NGSAO, with an average seeing of $\sim0.6''$.  The data was them binned and smoothed from its original resolution to $R\sim1000$ to increase signal to noise per resolution element. Details about the data reduction can be found in \cite{bowler_2014}.  In this study, we further binned the data to a resolution of $R$=300 for the retrievals. This allowed us to use the correlated-k opacities from \texttt{petitRADTRANS}, rather than substantially down-sampling the line-by-line opacities, which would result in inaccuracies, and reduced the run times for our retrievals. While medium resolution data has been suggested as a way to overcome the degeneracies of low-resolution data \citep[e.g.][]{hood_brown_2023}, we find that running our retrievals at R$\sim$1000 result in unphysical solutions due to low signal-to-noise and opacity sampling. These issues vanish when we bin to slightly lower resolution. We find no substantial differences to our retrieved parameters for a resolution of R$\sim$100-600, only a difference in run time. We opt for R$\sim$300 to optimize between resolution and run time.

\begin{deluxetable*}{clll}
\tablecaption{A summary table of all fitted parameters in our high-resolution (HR) and low-resolution (LR) \texttt{petitRADTRANS} retrievals and our adopted priors. The parameters for each of our two cloud models are detailed in separate sections. In the third column we note whether each parameter is used for our high-resolution (HR) or low-resolution (LR) retrievals. All parameters are used in our joint retrievals.\label{table:priors}}
\tabletypesize{\footnotesize}
\tablehead{
\colhead{Parameter} & \colhead{ Prior } & \colhead{Description} & \colhead{Retrieval}}
\startdata
R$_{P}$ & $\mathcal{U}$(1, 3) R$_{J}$ & planet radius in Jupiter radii & LR \\
log \textit{g} &  $\mathcal{U}$(2, 5.5) & log of the planet surface gravity & LR,HR\\
\textit{v} sin\textit{i} & $\mathcal{U}$(0, 20) km s$^{-1}$ & projected rotational velocity of the planet & HR \\
\textit{rv} &  $\mathcal{U}$(-10, 10) km s$^{-1}$  & apparent radial velocity of the planet & HR\\
 \textit{A$_v$} &  $\mathcal{U}$(0, 5) & optical extinction coefficient & LR \\
 \textit{b} & $\mathcal{U}$  & error inflation term & LR,HR\\
 scale K & $\mathcal{U}$(0.9, 1.1) & scale factor for $K$-band & LR\\ 
 scale H & $\mathcal{U}$(0.9, 1.1) & scale factor for $H$-band & LR \\ 
 \hline
 \multicolumn{3}{c}{Temperature Model Parameters}  \\
 \hline
$T_1,T_2,T_3$ & $\mathcal{U}$(0.1, 1) & anchor points for the spline pressure temperature profile (as fractions of adjacent points) & LR,HR\\
$\alpha$ & $\mathcal{U}$(1, 2) & power law for optical depth model & LR,HR \\ 
$\delta$ & $\mathcal{U}$(0, 1) & proportionality constant in optical depth model & LR,HR \\
$T_{int}$ & $\mathcal{U}$(300, 2700) K & interior temperature of planet & LR,HR \\
\hline
\multicolumn{3}{c}{Chemistry Model Parameters}  \\
\hline
C/O & $\mathcal{U}$(0.3, 1.5) &  C/O ratio of planet atmosphere & LR,HR \\
$[\text{Fe/H}]$ & $\mathcal{U}$(-1.5, 3) & log metallicity of planet atmosphere, relative to solar & LR,HR \\
P$_{\text{quench}}$ & $\mathcal{U}$(-6, 2) bars & carbon quench pressure for disequilibrium chemistry model & LR,HR \\
\hline
\multicolumn{3}{c}{Grey Cloud Model Parameters} \\
\hline
$\kappa_0$ & log-$\mathcal{U}$(-6, 10) & cloud opacity at the cloud base & LR,HR \\
$\gamma$ & $\mathcal{U}$(0, 15) & power law for scaling cloud opacity with pressure & LR,HR \\
$P_{0}$ & log-$\mathcal{U}$(-6, 2) bars & log pressure of the cloud base in bars & LR,HR  \\
\hline
\multicolumn{3}{c}{EddySed Cloud Model Parameters} \\
\hline
$\sigma_{\text{norm}}$ & $\mathcal{U}$(1.0, 3.0) & width of cloud particle size distribution & LR,HR\\
$f_{sed}$ & $\mathcal{U}$(0,15) & sedimentation efficiency (for calculating cloud deck parameters) & LR,HR\\
$K_{zz}$ & log-$\mathcal{U}$(5, 13) & vertical mixing efficiency (for calculating cloud deck parameters) & LR,HR \\
\textit{X$_{species}$} & log-$\mathcal{U}$(-4.5,1) & enhancement factor of given cloud species at the cloud base relative to equilibrium & LR,HR \\
& & abundance (species included: Fe, MgSiO$_3$, Al$_2$O$_3$.) &  \\
\enddata
\end{deluxetable*}

\subsection{High-resolution Spectroscopy of ROXs 42B b}

\cite{bryan_2018} obtained a high-resolution $K$-band spectrum of ROXs 42B b on UT June 1 and 2, 2015, using the pre-upgrade NIRSPEC instrument (R$\sim$25,000) on Keck II. The target was observed using a $0.041\times2.26$\arcsec~ slit in adaptive optics (AO) mode in order to minimize blending with the primary star. For ROXs 42B b, both the companion and primary (separation $\sim$ 1.2\arcsec) were observed simultaneously in the slit, with their light spatially separated on the detector. The host star's spectrum was then used to derive a wavelength solution and telluric correction for the lower signal-to-noise ratio (SNR) planetary data. 
Details on the observations and data reduction can be found in \cite{bryan_2018}. We used two spectral orders spanning $\lambda = 2.27-2.38$ $\mu$m in our retrieval because they contain absorption lines from both water and CO, including two strong CO bandheads.  The other spectral orders had less accurate telluric corrections and wavelength solutions, and we therefore follow the example of \cite{bryan_2018} and exclude them from our analysis.

In Figure \ref{fig:ccf} we show the cross-correlation function for our continuum-normalized, radial velocity-corrected high-resolution spectrum compared to single molecular template spectra. We detect H$_2$O and CO, which are expected to be the dominant near-infrared absorbers in an atmosphere of this temperature, at greater than 8$\sigma$. We do not detect CH$_4$ or OH, which are predicted to have much lower abundances.

\subsection{SNIFS Optical Spectrum of ROXs 42B AB}\label{sec:opt_spec}

In order to derive an estimate for the abundances of the binary host stars, we use a low-resolution optical spectrum obtained on UT May 20 2012 by the SuperNova Integral Field Spectrograph (SNIFS) at the University of Hawai'i 2.2-m telescope and published in \cite{bowler_2014}. SNIFS is an integral field unit with simultaneous coverage of the blue (3000-5200 \AA) and red (5200-9500 \AA) optical regions at a resolving power of R$\sim$1300. Additional details about the data reduction can be found in \cite{bowler_2014}. In this study, we  binned the SNIFs optical spectrum to a resolution of R$\sim$100 for comparison with the SPHINX model grid, which has a native resolution of R$\sim$250 \citep{iyer_sphinx_2023}.

\section{Atmospheric retrieval setup} \label{sec:retrieval_setup}

We use the retrieval framework implemented in \texttt{petitRADTRANS} \citep{molliere_petitradtrans_2019, molliere_2020} to model our low- and high-resolution spectra. We use a retrieval setup similar to the one described in \cite{molliere_2020}, which we describe in more detail below.

Following previous studies \citep[][etc.]{molliere_2020,burningham_2017,xuan_clear_2022}, we assume that the atmosphere of ROXs 42B b is in local chemical equilibrium, with the exception of potential quenched carbon chemistry. We use \texttt{petitRADTRANS} to calculate layer-dependent chemical abundances by interpolating tables generated by easyCHEM, a Gibbs Free Energy minimizer \citep{molliere_petitradtrans_2019}. We parameterize the bulk chemistry using a C/O ratio and metallicity, [Fe/H], which together set the abundances for all of the individual species. We perform initial fits to test for the potential presence of disequilibrium chemistry by fitting for a carbon quench pressure, which fixes the abundances of carbon monoxide, water and methane to a constant value at pressures lower than the quench pressure. We find that the quench pressure either tends to very low values or is completely unconstrained, and therefore exclude it from our final fits. This is not surprising, as methane is expected to have a very low abundance throughout ROXs 42B b's atmosphere and it is primarily the measured ratio of CH$_4$ to CO that constrains the quench pressure when the atmosphere is not in chemical equilibrium \citep{fortney_beyond_2020,xuan_clear_2022}.

\subsection{Opacities}\label{sec:opacities}

We use the pre-tabulated correlated-k and line-by-line opacities from \texttt{petitRADTRANS} \citep{molliere_petitradtrans_2019} to model our low and high-resolution spectra, respectively. We include opacities from CO, H$_2$O, FeH, CO$_2$, OH, H$_2$S, PH$_3$, Na, and K in our models. For completeness, we included the opacities of CH$_4$, HCN, and NH$_3$ in our initial fits, but found that this had no effect on our retrieved posteriors for either the low- or high-resolution fits. This is not surprising, as these molecules are all predicted to have very low abundances in the relatively hot atmosphere of ROXs 42B b. We find similar results for tests including VO and TiO. We find that the alkali species, Na and K, are particularly important for fitting our low-resolution data, as their absorption wings can affect the continuum shape at the blue edge of $J$-band. We use the same base line list from the VALD opacity database \citep{piskunov_vald_1995}, and try 3 different treatments of the wings, including the long wing treatment from \cite{burrows_calculations_2003}, custom pressure broadening \citep[for details see][]{molliere_petitradtrans_2019} and a Lorentz profile \citep{schweitzer_analysis_1996}, to quantify the effect of this choice on our retrieved abundances. We found that the differences were negligible and used the custom pressure broadening in our final fits. To cut down on computation time for our high-resolution retrievals, we down-sampled our opacities by a factor of 4 (native resolution of $10^6$, new resolution of $R$=250,000).  This new resolution is still 10$\times$ the resolution of our data.  We confirm that this resolution is sufficient for our data by generating synthetic data using \texttt{petitRADTRANS} and retrieve at both native and down sampled resolutions.  We find no significant differences in the retrieved posterior distributions.

\begin{figure*}[t!]
\epsscale{1.1}
\plotone{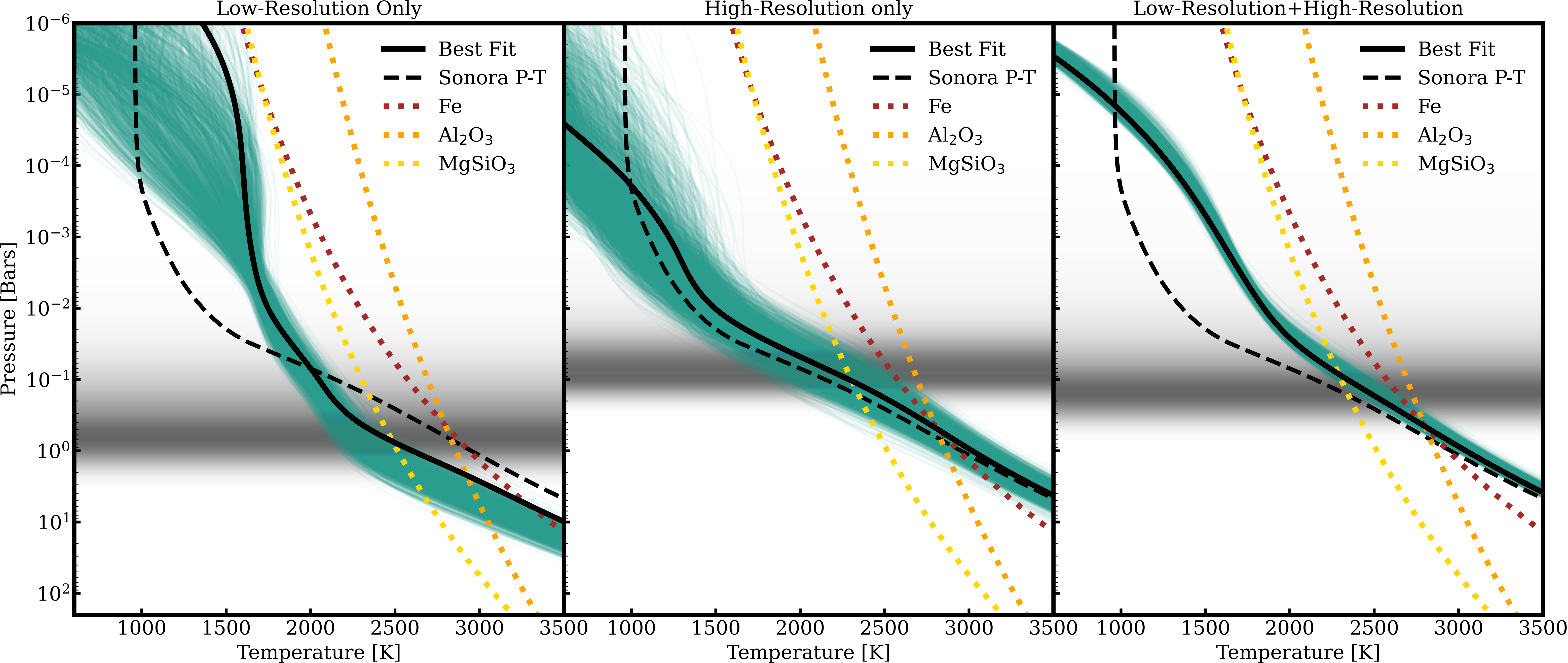}
\caption{Retrieved pressure-temperature (P-T) profiles for our best-fit models (bold in Table~\ref{table:bayes_factors}) for our low-resolution data (left), high-resolution data (center) and both combined (right). Randomly sampled P-T curves are plotted in green. Condensation curves for important cloud species in young hot objects, Fe, Al$_2$O$_3$, TiO$_2$ and MgSiO$_3$ are shown for comparison (dotted lines). We overplot a Sonora Bobcat model \citep{sonora_bobcat,marley_sonora_2021} corresponding to the photometrically derived temperature and log($g$). The grey gradient corresponds to the value of the wavelength-integrated flux contribution function.
\label{fig:PT_profiles}}
\end{figure*} 

\subsection{Temperature Structure} \label{PT_struc}
We adopt the 6-parameter spline pressure-temperature profile from \cite{molliere_2020}. This spline profile splits the atmosphere into three layers.  The lowest layer of the atmosphere is located beneath the radiative-convective boundary. In this region, the temperature profile is set equal to the moist adiabat corresponding to the assumed atmospheric temperature and composition.  The middle layer, which extends from the radiative-convective boundary to the point where the optical depth $\tau$ = 0.1, is a radiative photosphere where the temperature structure is set based on the optical depth and the classical Eddington approximation. The uppermost layer begins at $\tau$ = 0.1 and extends to a pressure of 10$^{-6}$ bars.  In this layer, the temperature is set using a cubic spline anchored by four equidistant points in log(P). This allows for a more complicated temperature structure than is permitted by the Eddington approximation, which forces the temperature at pressures above the photosphere to be isothermal. In our fits we penalize combinations of parameters that result in temperature inversions, which we do not expect to be present given ROXs 42B b's wide orbital separation.

\subsection{Clouds} \label{clouds}

Previous studies of isolated brown dwarfs and directly imaged planets with effective temperatures similar to that of ROXs 42B b have consistently found that clouds are required in order to accurately model their low-resolution spectra  \citep[e.g.][]{burningham_2017,burningham_2021,molliere_2020,lueber_2022}. Both \cite{bowler_2014} and \cite{currie_2014} found evidence for the presence of clouds in the atmosphere of ROXs 42B b when comparing to grid models. Spitzer mid-infrared spectroscopy of isolated brown dwarfs with spectral types similar to that of ROXs 42B b indicate that absorption from silicate grains is ubiquitous in these atmospheres \citep{suarez_ultracool_2023}. There are many cloud models of varying complexity in the literature, ranging from one parameter opaque cloud decks, to multiple cloud decks of different scale heights, pressure bases, and scattering properties. We consider two different cloud models in our retrievals: a simple, 3 parameter gray cloud deck, and the more complicated, physically motivated EddySed cloud model \citep{ackerman_2001, molliere_2020}.

For our gray cloud model, we consider a simple cloud deck with a wavelength-independent opacity, and parameterize the opacity by the cloud base pressure, P$_{base}$, the opacity at the cloud base, $\kappa_0$, and the pressure scaling power law, $\gamma$, as follows:
\begin{equation}
    \kappa(P) = \kappa_0 \bigg(\frac{P}{P_{base}}\bigg)^{\gamma}.
\end{equation}

In addition to this simple cloud prescription, we also carry out fits using the EddySed cloud model from \cite{ackerman_2001} as implemented in \texttt{petitRADTRANS} \citep{molliere_2020}. In this cloud model, the location of the cloud base is set by where the condensation curve of a given cloud species intersects with the atmospheric pressure temperature profile. The vertical extent and number density of the clouds is calculated using parameters representing the vertical mixing, $K_{zz}$, and the sedimentation efficiency $f_{sed}$. In addition, we fit for a multiplicative factor which scales the abundance of each cloud species relative to the equilibrium chemistry-predicted value at the cloud base, $X_{species}$.

We use a Mie scattering model to calculate the wavelength-dependent opacity from each cloud species in the EddySed model. We determine which cloud species to include in our model by comparing likely P-T profiles for ROXs 42B b to condensation curves for a wide range of potential cloud species. Based on this, we find that iron (Fe), enstatite (MgSiO$_3$), fosterite (Mg$_2$SiO$_4$), quartz (SiO$_2$) and corundum (Al$_2$O$_3$) are all possible cloud species that could condense in the atmosphere of ROXs 42B b. Signatures of silicates clouds have been observed in Spitzer IRS data of similar spectral type brown dwarfs \citep[e.g.][]{cushing_spitzer_2006,burgasser_clouds_2008,burningham_2021}. \cite{visscher_atmospheric_2010} predict that the formation of quartz clouds will be repressed in favour of enstatite due to the high abundance of Mg, though this has been shown to not always be true by \cite{burningham_2021}, whose retrievals favored enstatite clouds over deeper quartz clouds. On the other hand, \cite{gao_aerosol_2020} predict that fosterite should be the dominant silicate species to condense from equilibrium chemistry and cloud microphysics. However, the cross sections of quartz, fosterite, and enstatite clouds have a nearly identical slope at near-infrared wavelengths with no unique features to differentiate them \citep{wakeford_2015}, so we choose to focus on enstatite clouds for this study. We explore multiple combinations of cloud species including Al$_2$O$_3$, Fe, and MgSiO$_3$, as well as different cloud particle properties (amorphous vs. crystalline) for the solid species in our retrieval, which changes the refractive indices and resulting scattering cross sections. We find that our choice of amorphous versus crystalline structure has a negligible impact on our retrieved posterior probability distributions, so we use amorphous Mie scattering particles for our final models. This is consistent with model fits to mid-infrared Spitzer IRS spectra of brown dwarfs, which are best reproduced by amorphous silicate cloud particles  \citep{luna_empirically_2021,suarez_ultracool_2023}. 

\subsection{Additional Fit Parameters} \label{add_params}

We also fit for the surface gravity, parameterized as log($g$).  In the low-resolution retrievals, we additionally fit for a flux scaling parameter in order to match the overall flux level.  Our high-resolution spectrum is continuum normalized, and so is not sensitive to the absolute flux level. \texttt{petitRADTRANS} computes the flux density emitted at the surface of the planet. To obtain the observed flux, we must multiply this quantity by a scale factor:  
\begin{equation}
    f_{obs} = \bigg( \frac{R}{d} \bigg)^2 f_{emit}.
\end{equation}
where $R$ is the object's radius and $d$ is the distance to the system.

We allow for flux offsets between each of our low-resolution spectral bands (two offsets in total with $J$-band fixed) to account for the fact that they were observed with different instruments at different observational epochs. As this planet is located in a young star forming region, we additionally fit for an optical extinction coefficient, $A_V$, and correspondingly redden our model spectra to compare with our data. \cite{bowler_2014} calculated a reddening value for the primary binary of $1.7^{+0.9}_{-1.2}$  by comparing the optical SNIFS spectrum to optical templates from \cite{pickles_stellar_1998}. However, the constraints on this value are asymmetric and wide, and this value was calcuated assuming a single star, so we opt for a wide uniform prior on this value for our fits.

It is possible that the error bars of our data may be under-estimated due to the presence of correlated noise sources.  We therefore add a logarithmic error term, $b$, in quadrature to the reported measurement errors for each spectral bin, such that the effective standard deviation of each point, $i$, becomes:
\begin{equation}\label{eq:noise}
s_i^2 = \sigma_i^2 + 10^b.    
\end{equation}
For retrievals including the high-resolution data, we also include two additional parameters: the radial velocity offset, and the projected rotational velocity, $v$sin$i$. 

We adopt wide uniform or log-uniform priors for all parameters in our models and impose no restrictions other than excluding combinations of parameters that result in temperature inversions. A summary table of all the parameters included in the retrieval and their priors is shown in Table~\ref{table:priors}. 

\begin{deluxetable*}{cccccccc}
\tablecaption{Best-fit values and confidence intervals for key parameters from ROXs 42B b retrievals. 
\label{table:bayes_factors}}
\tabletypesize{\footnotesize}
\tablehead{Cloud Model & C/O & [Fe/H] & Radius ($R_{Jup}$) & $v$sin$i$ (kms$^{-1}$) & log($g$) (cgs) & $T_\textrm{eff}$ (K) & Bayes Factor ($\sigma$)
}
\startdata
 \multicolumn{8}{c}{High-Resolution Retrievals} \\ \hline
 \textbf{Clear} & $0.504\pm0.048$ & $-0.67\pm0.35$ & N/A & $10.52\pm0.92$ & $3.49\pm0.57$ & $2720\pm80$ & 1.0 \\
 Grey Clouds & $0.510\pm0.047$ & $-0.64\pm0.31$ & N/A & $10.57\pm0.83$ & $3.46\pm0.48$ & $2800\pm100$ & 0.18 ($-2.4\sigma$) \\
 EddySed (MgSiO$_3$ + Fe)  & $0.510\pm0.032$ & $-0.69\pm0.33$ & N/A & $10.35\pm0.65$ & $3.40\pm0.45$ & $2750\pm80$ & $2.8\times10^{-8}$ ($-6.2\sigma$)\\
 \hline
 \multicolumn{8}{c}{Low-Resolution Retrievals} \\
 \hline
 \textbf{Clear} & $0.723\pm0.029$ & $1.37\pm0.13$ & $2.74\pm0.06$ & N/A & $3.9\pm0.2$ & $1890\pm35$ & 1.0 \\
 Grey clouds & $0.705\pm0.031$ & $1.33\pm0.17$ & $2.78\pm0.06$ & N/A & $3.70\pm0.21$ & $1886\pm36$ & 0.15 ($-2.5\sigma$)\\
 Grey clouds + fixed P-T & $0.724 \pm0.007$ & $1.37\pm0.08$ & $2.72\pm0.05$ & N/A & $3.58\pm0.11$ & $1876\pm27$ & 0.37 (-2.0$\sigma$) \\
 Grey clouds + fixed P-T, $A_V$ & $0.732\pm0.007$  & $1.11\pm0.08$ & $2.61\pm0.03$ & N/A & $3.31\pm0.06$ & $1834\pm11$ & $2.1\times10^{-9}$ (-6.6$\sigma$) \\
 EddySed (MgSiO$_3$) & $0.704\pm0.033$ & $1.35\pm0.14$ & $2.77\pm0.06$ & N/A & $3.79\pm0.19$ & $1895\pm39$ & 0.37 ($-2.0\sigma$) \\
 EddySed (Fe, am) & $0.694\pm0.035$ & $1.33\pm0.18$ & $2.78\pm0.06$ & N/A & $3.78\pm0.20$ & $1894\pm34$ & 0.67 ($-1.6\sigma$) \\
 EddySed (MgSiO$_3$ + Fe, am) & $0.696\pm0.038$ & $1.34\pm0.16$ & $2.77\pm0.06$ & N/A & $3.81\pm0.20$ & $1898\pm40$ & 0.034 ($-3.1\sigma$)\\
 EddySed (MgSiO$_3$ + AlO$_2$, am) & $0.702\pm0.032$ & $1.34\pm0.16$ & $2.78\pm0.05$ & N/A & $3.74\pm0.18$ & $1893\pm38$ & 0.54 ($-1.8\sigma$)\\
\hline
\multicolumn{8}{c}{Joint Retrievals}  \\
\hline
 Clear & $0.778\pm0.003$ & $0.37\pm0.03$ & $2.84\pm0.02$ & $5.55\pm0.74$ & $2.76\pm0.08$ & $1843\pm22$  & 1.0 \\
 \textbf{Grey clouds} & $0.734\pm0.008$ & $0.46\pm0.04$ & $2.83\pm0.01$ & $5.57\pm0.35$ & $2.94\pm0.05$ & $1935\pm12$ & $1808$ ($4.3\sigma$)\\
 EddySed (MgSiO$_3$ + Fe)  & $0.691\pm0.004$ & $0.67\pm0.02$ & $2.81\pm0.01$ & $10.30\pm0.35$ & $3.47\pm0.017$ & $1905\pm17$  & $44.7$ ($3.2\sigma$) \\
\enddata
\tablecomments{The right most column lists the Bayes factor ($B$) for each retrieval relative to the clear model ($B=1$). The uncertainties for each parameter quoted are determined using the 68\% confidence interval from our posteriors assuming a Gaussian distribution. For our cloudy models, `am' stands for amorphous cloud particles + Mie scattering. Our reported Bayes factor are calculated relative to the clear model and then converted to a minimum sigma value using the method described in \cite{benneke_how_2013}}
\end{deluxetable*}

\subsection{Nested Sampling}

We fit the data using nested sampling as implemented in PyMultiNest, a python wrapper for MultiNest \citep{feroz_2009}. We run our retrievals with 1000 live points, and perform multiple runs using the optimal sampling efficiency for PyMultiNest to compute our evidence and posterior separate. Our ``best fit'' model is taken to be the sample with the highest likelihood. We compare different models using the Bayes factor, B$_{12}$, which gives the evidence for model 1 relative to model 2. It is calculated as:
\begin{equation}
    B_{12} = \frac{Z_1}{Z_2},
\end{equation}
where Z is the evidence for each given model as computed by PyMultiNest.  The higher the Bayes factor, the stronger the evidence supporting one model over another. We evaluate the Bayes factor for each of our different models relative to our best fit model and list the results in Table~\ref{table:bayes_factors}.

\begin{figure*}[t!]
\epsscale{1.1}
\plotone{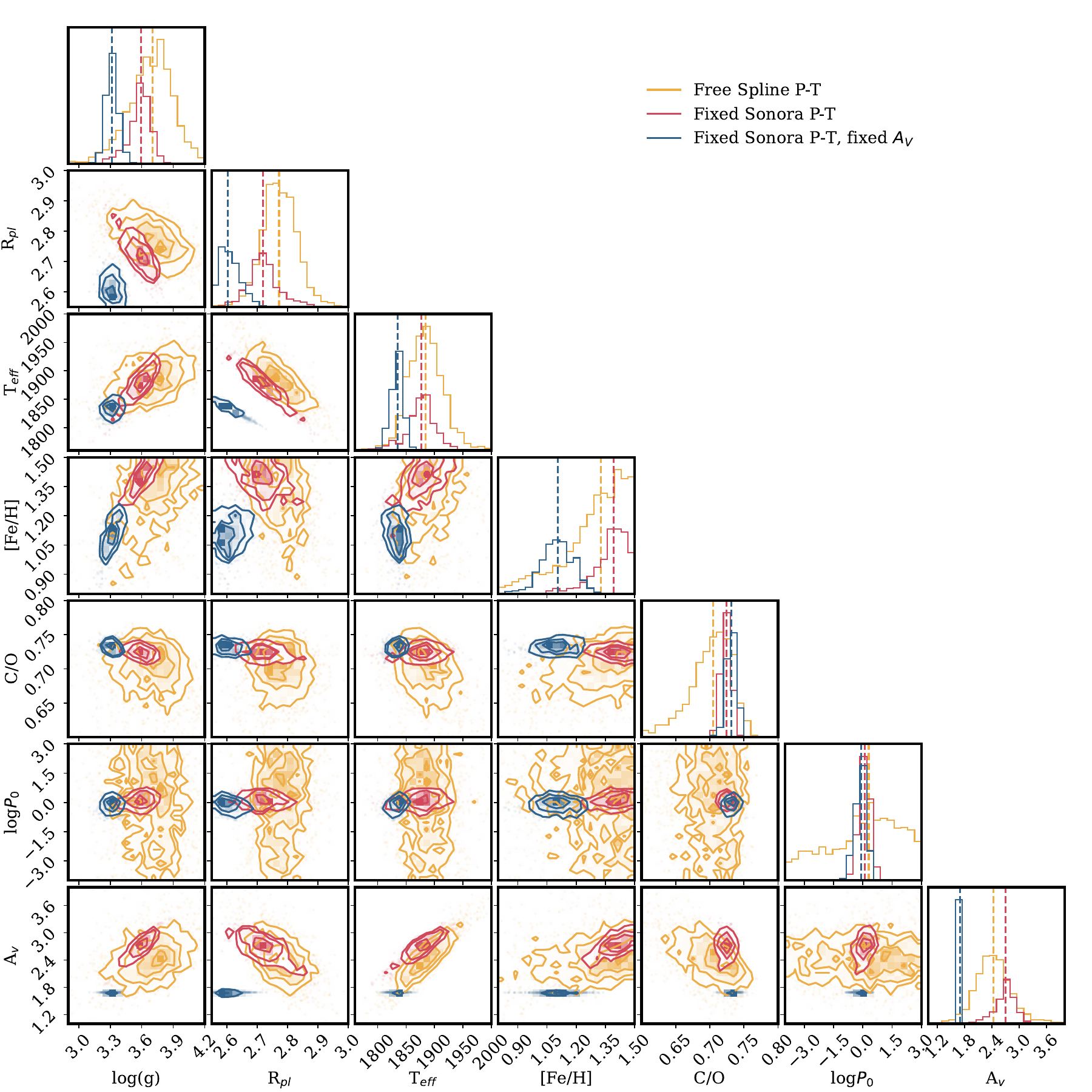}
\caption{Corner plot comparing the posterior probability distributions of key parameters from our fixed and free P-T profile retrievals on our low-resolution spectrum using gray cloud model and a) unrestricted spline pressure-temperature (P-T) profile, b) the P-T profile fixed to a Sonora P-T profile \citep{sonora_bobcat,marley_sonora_2021}, and c) a P-T profile fixed to a Sonora model P-T profile and the reddening coefficient fixed to the best fit stellar value from \cite{bowler_2014}. 
\label{fig:PT_comp_corner}}
\end{figure*} 

\section{Atmospheric Retrievals} \label{sec:retrievals_all}

\begin{figure*}[t!]
\epsscale{0.95}
\plotone{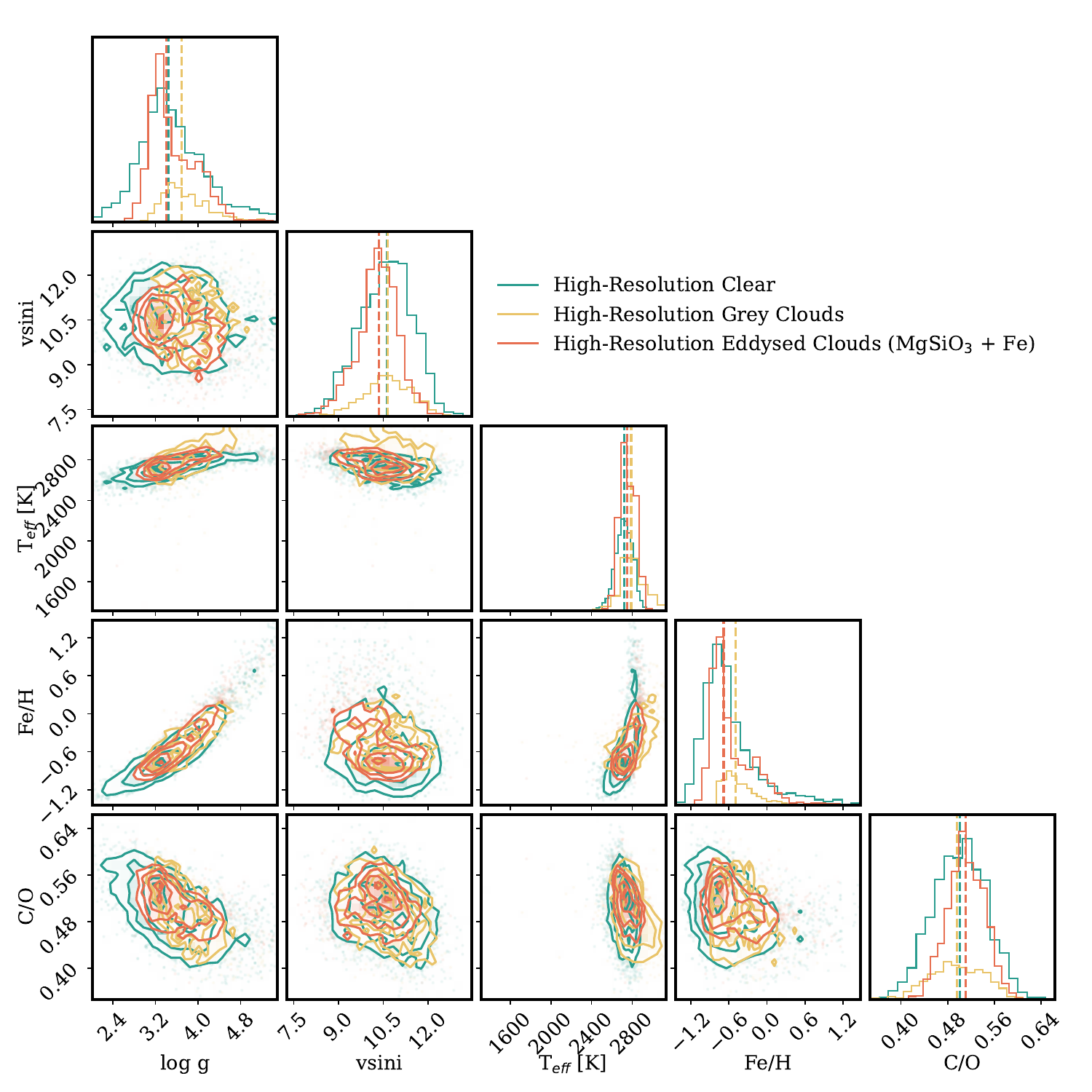}
\caption{Corner plot comparing the posterior probability distributions of key parameters from retrievals on our high-resolution spectrum for three different cloud models: a clear sky, a simple gray cloud deck, and EddySed condensate clouds composed of Fe and MgSiO$_3$. We find identical posterior probability distributions regardless of the cloud model used. 
\label{fig:highres_corner}}
\end{figure*}  

\subsection{Low-Resolution Retrievals} \label{sec:Low_res}

In this section we describe the results of our retrievals on our low-resolution spectra. We fit our low-resolution data using 5 different cloud models: a `clear' model, where there are no clouds present in the visible portion of the atmosphere, a gray cloud model, and 3 variations of the EddySed cloud model to account for the unknown cloud properties. Our variations of the EddySed cloud model include  an iron cloud deck, an enstatite cloud deck, and a combination of iron and silicate clouds, as well as a combination of enstatite and corundum clouds. In Figure~\ref{fig:LR_spec} we show the best fit spectra and residuals for our clear model, the best fitting model. In Figure~\ref{fig:PT_comp_corner} we show the posteriors for a few select parameters from each retrieval, including log($g$), $T_{eff}$, C/O ratio and [Fe/H]. The best fit values and 1-sigma uncertainties for key parameters are reported in Table~\ref{table:bayes_factors}.

\begin{figure*}[h]
\centering
\epsscale{1.1}
\includegraphics[width=\linewidth]{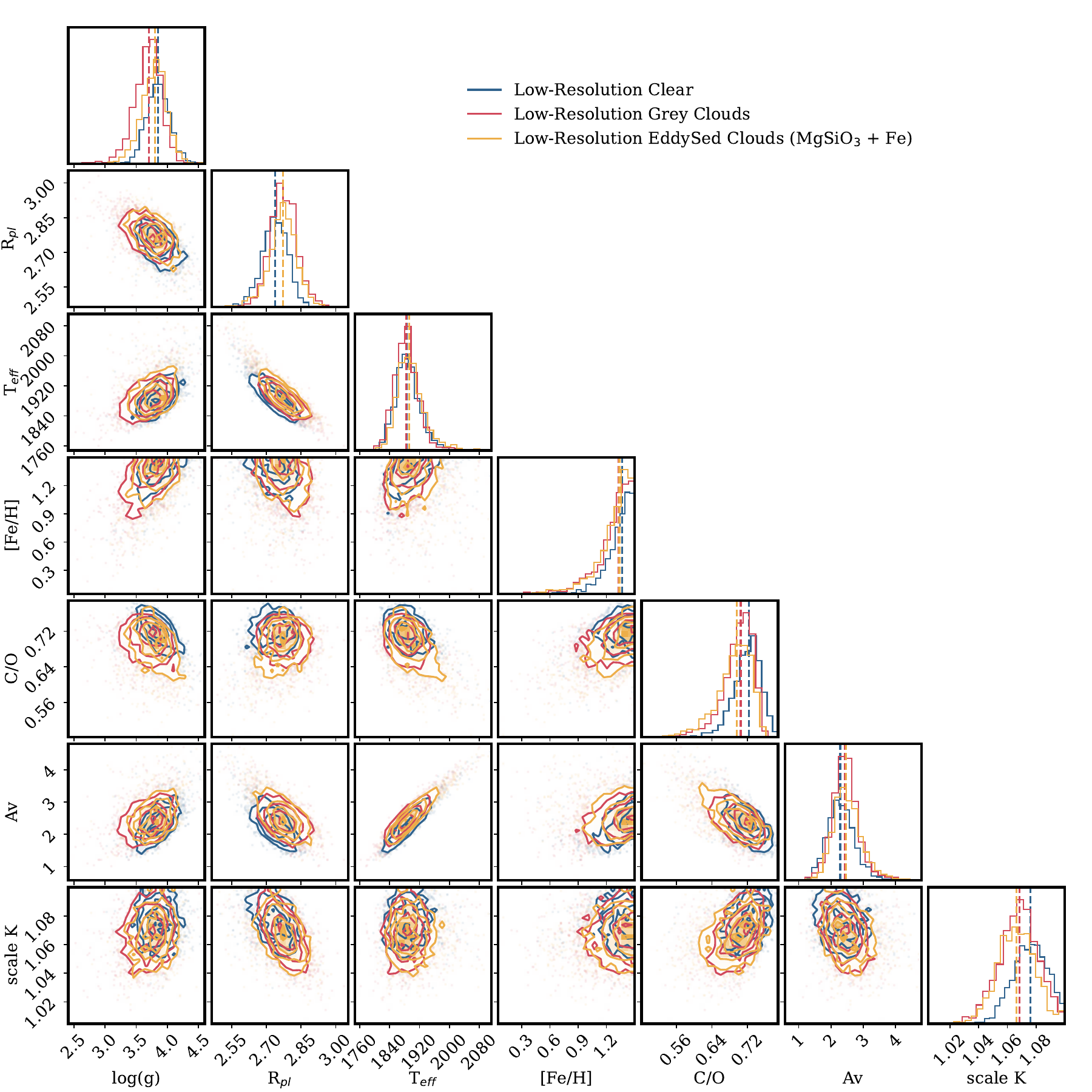}
\caption{Corner plot comparing the posterior probability distributions of key parameters from retrievals on our low-resolution spectrum for our three considered cloud models, a clear sky, a simple gray cloud deck, and EddySed condensate clouds composed of Fe and MgSiO$_3$.
\label{fig:lr_corner}}
\end{figure*}

We find that our clear model is preferred at 2.4 sigma over our grey cloud model, and 6.2 sigma from our EddySed cloud model. We find a super-solar metallicity of $1.37\pm0.13$. We also find a super-solar C/O ratio of $0.798\pm0.019$. Our retrieved C/O ratio, [Fe/H] and P-T profile are consistent between our clear, grey and EddySed cloud models (Fig.~\ref{fig:lr_corner}). When cloud parameters are included in our fits, we find either clouds too deep in the atmosphere to affect the spectrum, or completely unconstrained values, which indicates that clouds are not necessary to reproduce our data or there is insufficient information to constrain cloud properties.

Even with constraints on the anchor points of our spline P-T profile, our fits strongly prefer an isothermal temperature structure at pressures lower than 1 bar. These kinds of steep, isothermal P-T profiles have been seen in other studies \citep[e.g.][]{burningham_2017} and are thought to be an attempt to compensate for the effect of cloud opacity using the flexibility of the P-T profile. While it has been suggested by \cite{tremblin_cloudless_2017} that the atmospheres of these planets could in fact be isothermal and the observed reddening attributed to clouds could be explained by fingering convection, we find this is unlikely in the case of ROXs 42B b. Given that these objects are internally heated, it is very unlikely that their atmospheres would be isothermal; indeed, radiative equilibrium grid models predict a much steeper temperature profile for ROXs 42B b \citep[e.g. the ATMO and Sonora Bobcat model grids.][]{phillips_new_2020,sonora_bobcat,marley_sonora_2021}. The changing colors of brown dwarfs as a function of effective temperature, as well as the high frequency of rotational variability near the L-T transition, all provide strong evidence that clouds should be ubiquitous in the atmospheres of objects like ROXs 42B b \citep{bailer-jones_search_1999,gelino_l_2002,marley_patchy_2010,apai_hst_2013,radigan_strong_2014,vos_search_2019}.

\begin{figure*}[t!]
\epsscale{1.1}
\plotone{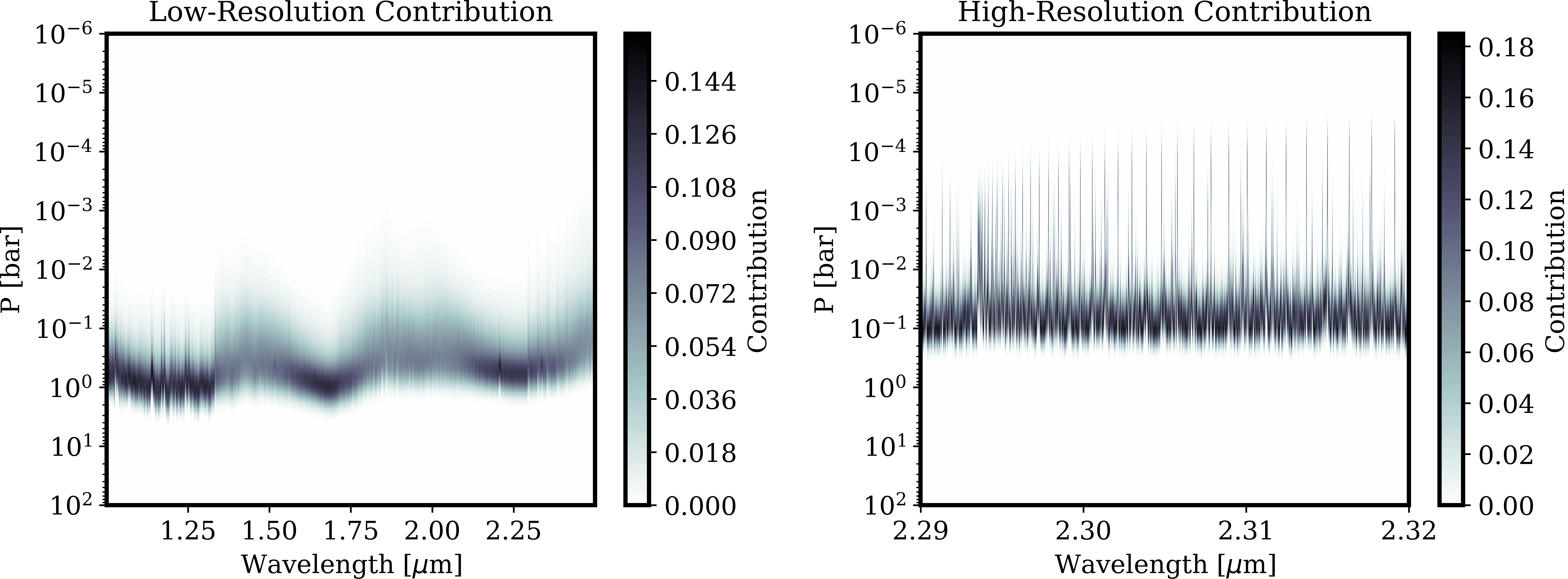}
\caption{Contribution functions for the best-fit models for our low-resolution (left) and high-resolution (right) retrievals. 
\label{fig:contribution_func}}
\end{figure*} 

We also consider fits where we fix the P-T profile to a self-consistent Sonora model \citep{sonora_bobcat,marley_sonora_2021} with log($g$) = 3.5 and an effective temperature of 2100 K, matching the spectroscopically and photometrically derived parameters for ROXs 42B b \citep{currie_direct_2013,currie_2014}. With this fixed P-T profile, we find that we are no longer able to adequately fit our low-resolution data using a cloud free model, as there are no combination of parameters that can reproduce the data. With the P-T profile fixed, we find a smaller radius and $T_{eff}$ than in our free P-T retrieval case, indicating that is trading off with the radius in an attempt to conserve total luminosity. We additionally find a higher reddening coefficient, indicating the need for extra absorption to match our observed spectrum. 

When we include gray clouds in our fixed P-T profile retrievals, we are able to match our low-resolution spectrum nearly as well as the retrievals with the free P-T profile, with the free P-T profile model preferred only slightly at 2$\sigma$. We find that the cloud base becomes constrained in the retrievals, but the optical depth of the clouds trades off with the reddening coefficient. We therefore try an additional fit where we fix the reddening value to the value calculated by \cite{bowler_2014} for the primary binary. In Figure~\ref{fig:PT_comp_corner} we compare the posteriors for our fixed P-T profile retrievals to our free P-T profile retrieval using the gray cloud model. We find that with the reddening and P-T profile fixed, cloud opacity becomes important to match our observed spectrum, and the metallicity, log($g$) and planetary radius values all decrease substantially, while the C/O ratio increases only slightly. We conclude that there is a degeneracy between our retrieved chemistry and atmospheric temperature structure. It is most strongly observed to effect our retrieved metallicity. We find that our retrieved metallicity increases or decreases as the slope of our P-T profile changes from more isothermal to less. This makes it evident that the model is using the effects of the P-T profile to account for the effects of cloud on absorption depth, as a more isothermal atmospheres makes it harder to have deep absorption features. The metallicity as a result is driven up to account for the observed absorption features.

To confirm this, we perform similar tests restricting our metallicity to values lower than 0.5. In this case, the P-T structure is less isothermal. We observe the worst fit to occur in the $H$-band peak, suggesting this driving this degeneracy (Fig.~\ref{fig:LR_spec}). Since this feature is also a strong indicator of gravity, we therefore conclude that we cannot trust our retrieved gravity value either. We additionally observe a degeneracy between our optical extinction coefficient, $A_V$, and the cloud properties. For our free P-T profile retrievals, $A_V=  2.43\pm0.41$, and for our fixed P-T profile retrievals, it increases slightly to $2.67\pm0.25$. We find that the retrievals prefer to increase the $A_V$ value rather than add clouds, likely because this involves toggling only a single parameter verses a combination of parameters, as the net effect of silicate clouds is to redden the observed spectrum.

In all versions of the fit, we consistently struggle to fit the $J$-band data with our models, particularly at the blue edge. This may be related to the extended wings of the Na and K lines. We compare retrievals using three different wing treatments described in \S\ref{sec:opacities}, but are still unable to replicate the observed $J$-band shape. TiO also has absorption features around 1.2 microns, however we find that even when we include TiO with free abundance, it does not improve the quality of our fit. 

Based on the observed degeneracies between various parameters in our low-resolution retrievals, we conclude that our retrieved abundances and other properties are likely unreliable.

\subsection{High-resolution Retrievals} \label{sec:High_res}

In this section, we present the results of our retrievals on the high-resolution data and compare with the results from our retrievals using the low-resolution data. We run the same suite of cloud models as we did on our low-resolution data and additionally fit for a projected rotational velocity ($v$sin$i$) and a radial velocity (RV) offset. The best-fit values and uncertainties for key parameters including the C/O ratio, [Fe/H], $v$sin$i$, log($g$) and $T_{eff}$ are listed in Table~\ref{table:bayes_factors}. 
  
For our high-resolution data, we also find no difference in the quality of the fit between models with and without clouds. We compare the posteriors for the clear, grey cloud, and EddySed cloud models in Figure \ref{fig:highres_corner} and find that they all overlap. As a result, the Bayes factor prefers the simpler cloud-free model. Since our inferred parameters are not sensitive to our choice of cloud model, we use the parameters from the cloud-free model moving forward. We retrieve a moderately substellar metallicity of [Fe/H] = $-0.67\pm0.35$, rather than the super-stellar metallicity preferred in our low-resolution retrievals.  We find a C/O ratio of $0.505\pm0.05$, which is consistent with the solar C/O ratio of $0.59\pm0.13$ \citep{asplund_chemical_2021}. We find a radial velocity value $rv=1.44\pm0.42$ kms$^{-1}$. We additionally constrain the projected spin rate $v$sin$i$ to be $10.52\pm0.92$ km s$^{-1}$, which is in good agreement with the value of $9.5^{+2.1}_{-2.3}$ km s$^{-1}$ found by \cite{bryan_2018} for these same data using a radiative equilibrium model and calculating the autocorrelation function. It is important to note that we utilized the instrumental broadening value reported by \cite{bryan_2018} in our fits rather than re-deriving our own estimate of this parameter; this means that our result is not a fully independent confirmation of the measurement reported in this study. 

Unlike with the low-resolution data, the P-T profile in the cloud-free retrieval does not appear to be compensating for the effects of clouds by adopting an isothermal stratosphere. In Fig.~\ref{fig:PT_profiles} we show that the shape of our derived P-T profile for cloud free retrieval agrees with predictions from the cloudless Sonora Bobcat forward model grid for an object of the same effective temperature and log($g$) \citep{sonora_bobcat,marley_sonora_2021}, though it is possible that the presence of clouds would alter this profile. For models with clouds included, we find that the cloud parameters are unconstrained. This indicates that our high-resolution data are not biased in the same way as the low-resolution retrievals by the presence of clouds, and are instead relatively unaffected by the presence of clouds.

In order to investigate why our high-resolution retrievals do not appear to be biased by the presence of clouds the same way our low-resolution retrievals are, we examine where the flux originates in our atmosphere relative to the expected positions of cloud layers. In Fig.~\ref{fig:contribution_func} we plot the flux contribution functions for our best-fit models for our high-resolution and low-resolution retrievals. In the low-resolution retrievals, we see that the flux primarily originals at pressures greater than 10$^{-1}$ bars, with some smearing out to lower pressures in the strong water features. For our high-resolution data, a substantial portion of the flux is emitted higher in the atmosphere, in the cores of individual lines. In Fig.~\ref{fig:PT_profiles}, we show the contribution compared to our P-T profiles and condensation curves. We see that the intersection between the P-T profiles and condensation curves occurs deeper than where the flux contribution peaks in our high-resolution data.  Based on this, is seems that the high-resolution data is less sensitive to clouds because the cores of the absorption lines probe higher in the atmosphere, above the clouds. In other words, the wavelength-averaged contribution function for our high resolution data lies above the predicted cloud base for all three cloud species considered here (see Fig.~\ref{fig:PT_profiles}). While the continuum is probably still affected by clouds, we lose continuum information in our data analysis. Larger wavelength coverage out to other bands, or flux calibrated high-resolution data should improve the sensitivity of high-resolution data to cloud properties.

\subsection{Joint Retrievals} \label{sec:Joint_Ret}

\begin{figure*}[h]
\centering
\epsscale{1.1}
\includegraphics[width=\linewidth]{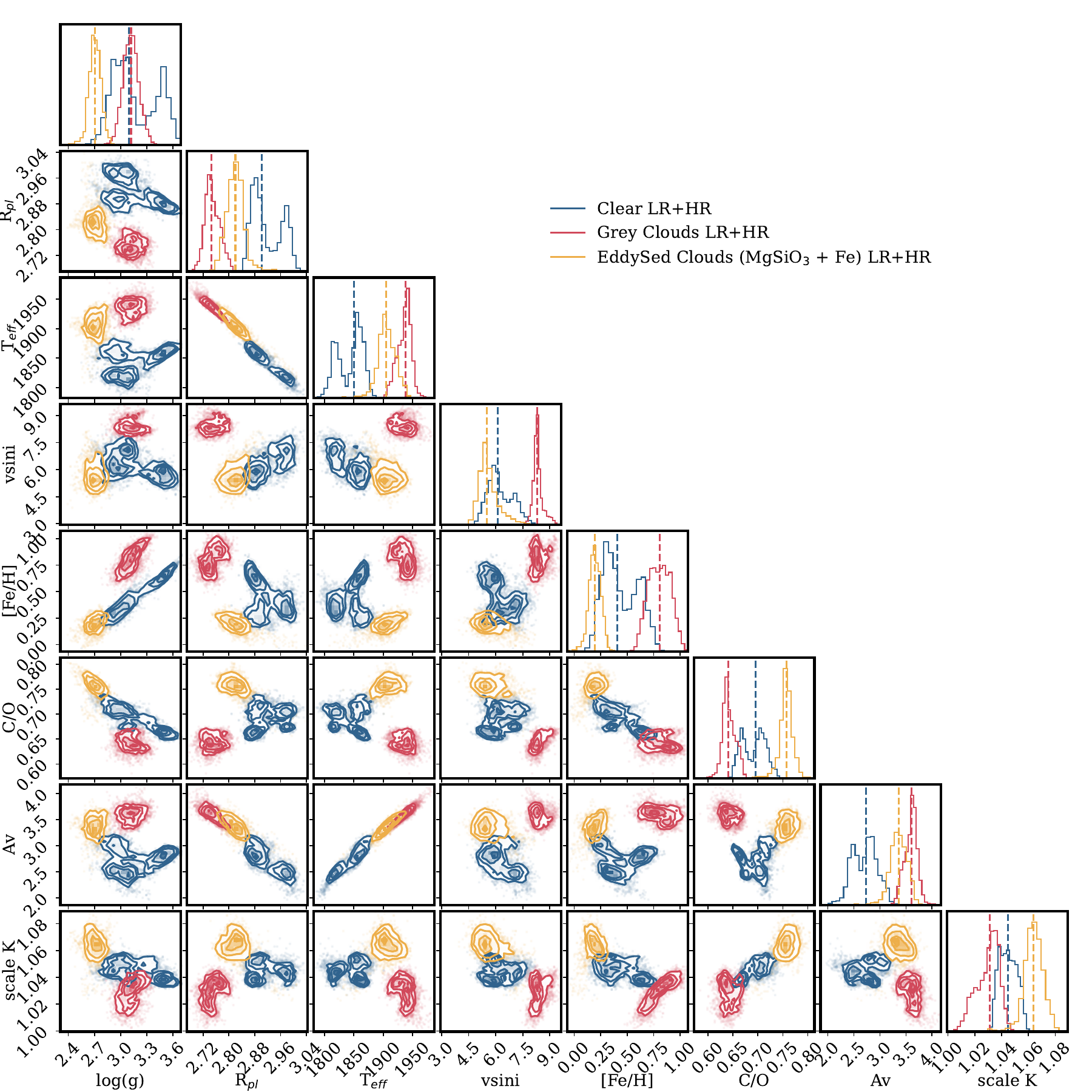}
\caption{Corner plot comparing the posterior probability distributions of key parameters from joint retrievals on our low-resolution and high-resolution data sets for our three considered cloud models, a clear sky, a simple gray cloud deck, and EddySed condensate clouds composed of Fe and MgSiO$_3$.
\label{fig:joint_corner}}
\end{figure*}

In addition to fitting both our high-resolution and low-resolution data sets for ROXs 42B b individually, we fit both together in a joint retrieval. As with the individual retrievals, we consider all three types of cloud model, and fit for all the parameters shown in Table~\ref{table:priors}. We show a subset of our retrieved values for these retrievals in Table~\ref{table:bayes_factors} and the full posteriors are shown in Fig~\ref{fig:joint_corner}.

In this joint fit, we find a lower metallicity and better constrained cloud properties for both our gray clouds and EddySed models, which were unconstrained in the fits using only the low-resolution data. We additionally find that our retrieved parameters depend strongly on the cloud model used. Like in other retrievals, we observe a trade-off between retrieved radius and $T_{eff}$, which is expected as those parameters together determine the total luminosity of the planet. For our cloud-free case, we find a multi-modal posterior, and that it is disfavored over either of our cloudy models.

We find that the addition of the high-resolution data forces the model P-T profiles to be less isothermal than the low-resolution data alone. As discussed in \S\ref{sec:Low_res}, it is highly unlikely for clouds not to be present in the atmosphere of a young hot object like ROXs 42B b, and more likely degeneracies in our models that resulted in our low-resolution retrievals favoring cloud free models. As a result, the Bayes factor strongly favors cloudy models, as cloud free models cannot provide a good match to both the shape of the low- and high-resolution spectra.  Likely because it has fewer degrees of freedom and due to the lack of strong cloud features in the considered wavelength range, we find that our grey cloud model is preferred from the Bayes factor. 

\begin{figure}[t!]
\epsscale{1.2}
\plotone{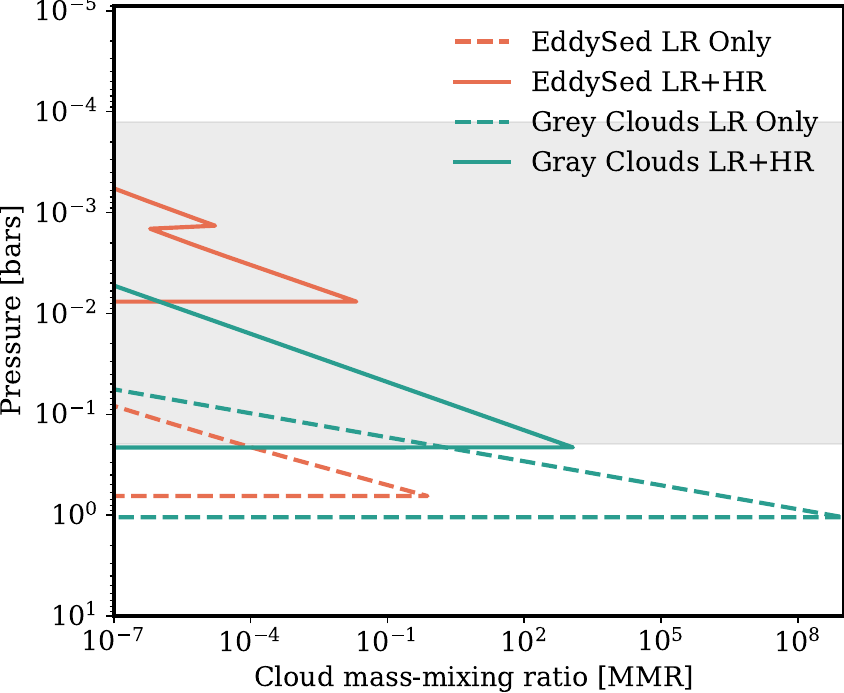}
\caption{The retrieved cloud opacity for our EddySed cloud model (orange), compared with the retrieved cloud opacity for our grey slab cloud model (green) as a function of pressure, for our low-resolution only fits (dashed lines) and our joint high-resolution and low-resolution fits (solid lines). The grey shaded region represents the region of the atmosphere where the total contribution to flux is greater than 0.01$\%$. For both cloud models we see that the in the low-resolution only case, the clouds are placed deep in the atmosphere where they have minimal impact on the flux, and the inclusion of the high-resolution data requires the presence of clouds to explain the observed data.
\label{fig:clouds}}
\end{figure}

In the third panel of Fig.~\ref{fig:PT_profiles}, we show our retrieved P-T profile for our preferred model. We find our P-T profile is consistent in the deep atmosphere with our high-resolution retrievals, but prefers a hotter upper atmosphere than the high-resolution retrievals alone. Our retrieved value for \textit{v}sin\textit{i} is also sensitive to our choice of cloud model in the joint fits. This is concerning, as the low resolution data do not contain any additional information about this parameter.  For our EddySed cloud model, our retrieved value agrees with our high-resolution only retrievals and the value from \cite{bryan_2018}, while the clear and grey cloud models find a lower value. 
This suggests that the need to match both data sets simultaneously alters the preferred solution in a way that biases our measurement of \textit{v}sin\textit{i} away from the true value.

\begin{figure*}[t!]
\centering
\includegraphics[width=0.8\linewidth]{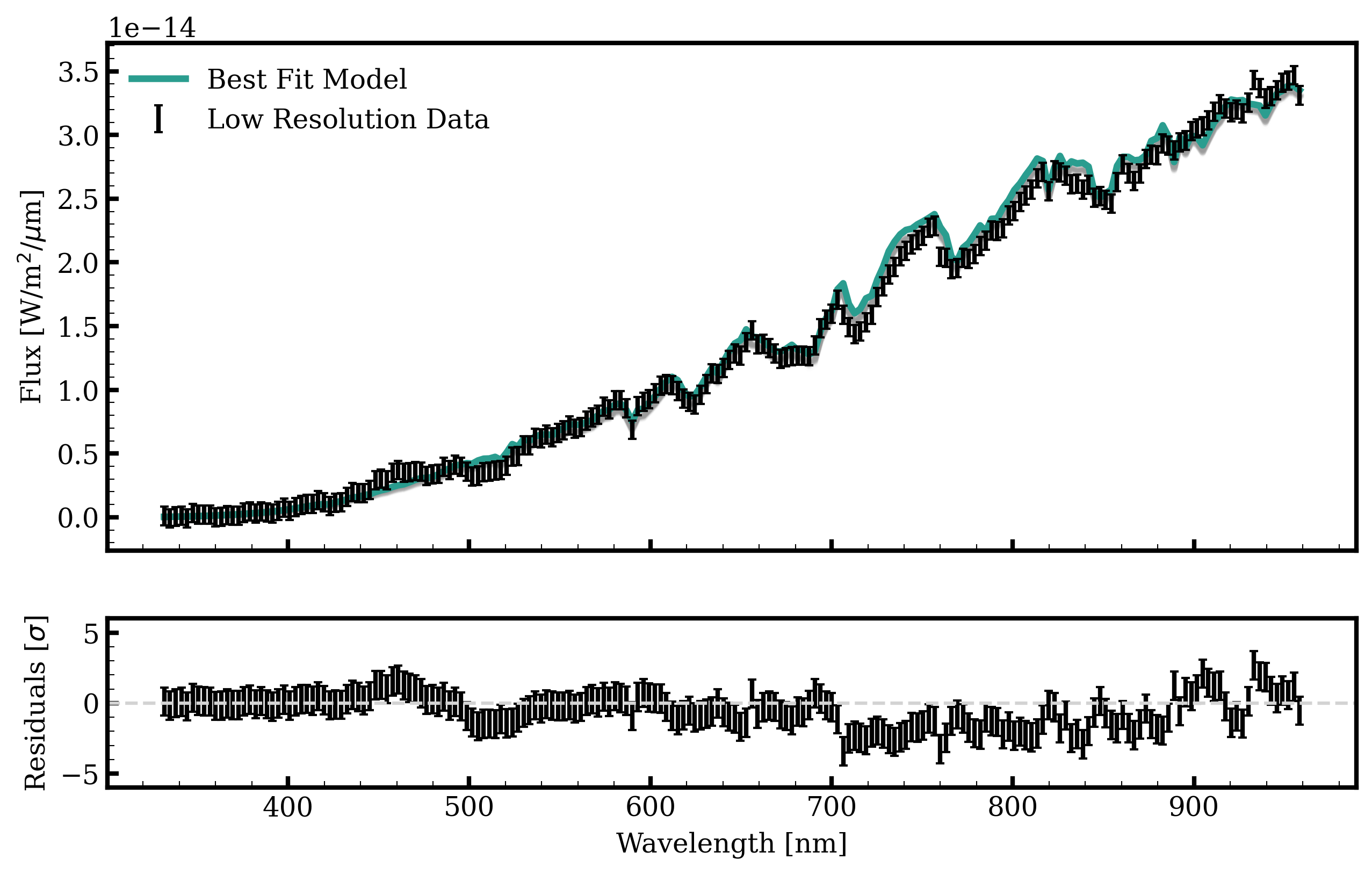}
\caption{Best fit model to the SNIFS optical spectrum of the unresolved binary ROXs 42B (top panel) and error-weighted residuals (lower panel). The best fit model is shown in green, with random drawn samples shown in gray.
\label{fig:starspec}}
\end{figure*} 

In Fig.~\ref{fig:clouds} we compared our retrieved cloud opacity for our retrievals on our low-resolution data only (dashed lines), and our joint retrievals (solid lines) for both our grey cloud models (green) and our EddySed cloud models (orange). We find that when we fit just the low-resolution data, the clouds are forced deep into the atmosphere where they do not affect the spectrum, while the P-T profile becomes isothermal to compensate for their effect (Fig.~\ref{fig:PT_profiles}). When we jointly fit the low and high-resolution data, clouds are required to produce the observed absorption features of the high-resolution data and the shape of the low-resolution spectrum, forcing them into the visible region of the atmosphere.

Given the strong dependence of cloud model on our retrieved abundances and planet properties, we decide to adopt the parameters from our high-resolution-only retrieval for subsequent analysis.

\section{Host Star Abundances} \label{sec:hoststar}

We fit for host star abundances by performing a `grid-retrieval' where we interpolate the SPHINX model grid from 
\cite{iyer_sphinx_2023}. The SPHINX model grid spans the relevant parameter space for M-type stars in $T_{eff}$ ($2000-4000$ K), gravity (log($g$) between $4-5.5$), C/O ($0.3-0.9$), and metallicity (log($Z$) between $-1.0-1.0$). The SNIFS optical spectrum (see \S\ref{sec:opt_spec}) contains the blended light from both binary components, so we model the spectrum as the sum of the flux contributions from each binary member. We then fit for the stellar radius, $R_*$, effective temperature, $T_{eff}$ and surface gravity, log($g$) of each binary component. At each step in the fit, we scale our models to our flux calibrated spectrum using the Gaia measured distance of our system \citep[146.447][]{gaia_collaboration_vizier_2022} and our stellar radii (see Table \ref{table:systemvals}). We place a uniform prior of $0.5-3.5$ on the binary flux ratio based on the $K$-band ratio reported by \cite{kraus_2013}.  

We assume that both components of the binary share a single C/O ratio and bulk metallicity. Published studies have found that the metallicities of wide separation binaries can differ by up to $0.03-0.1$ dex, but this decreases with decreasing binary separation \citep[e.g.][]{hawkins_2020,behmard_planet_2023}.  In this case, the binary components are only separated by $\sim$10 au (83 mas) \citep{kraus_2013}, and any differences in composition are therefore likely to be smaller than our measurement errors.  We additionally fit for an optical extinction coefficient, $A_V$, to account for extinction from dust and gas in the star forming complex. 
We also include an error inflation term as defined in Eq.~\ref{eq:noise} to allow for the possibility of data-model mismatches. \cite{iyer_sphinx_2023} found that their models differed from the measured spectral shapes more at shorter wavelengths, suggesting that this region may be particularly challenging from a model fitting perspective. We use nested sampling with 1000 initial live points. 

We list the retrieved values for each component of the binary in Table~\ref{table:systemvals}. We obtain an optical extinction coefficient of $A_V$ = $1.4\pm0.1$, which is consistent with the value reported by \cite{bowler_2014}. Our retrieved stellar radii are $1.42\pm0.01$ R$_{\odot}$ and $1.59\pm0.02$ R$_{\odot}$. Both of these values are consistent with predictions from the BHAC15 evolutionary model grids for low mass stars at the approximate age of these system \citep{baraffe_new_2015} using the measured masses of the binary components from \cite{kraus_2013}. However, when we invert our posteriors for our log($g$) and radius, we find mass values of: M$_1$ = $1.12\pm0.13$ M$_{\odot}$ and
 M$_2$ = $1.29\pm0.26$ M$_{\odot}$. While the mass of the more massive component is consistent within 2 $\sigma$, the less massive component is inconsistent at 3.5$\sigma$. This could be potentially due to flux calibration errors, a slightly incorrect distance, or slight issues with the models leading to an incorrect radius. Alternatively, since this is a very young system, activity of the stars could be contributing. 

 We note the disagreement between our models and data from 7000 - 9000 nm. This could be due to contamination from the strong telluric bands present at these wavelengths, missing opacity in the model, or interpolation errors from the limited grid resolution. Regardless, we overall find a decently good fit to our data, with values that match predictions for this system.

\begin{figure}[t!]
\epsscale{1.2}
\plotone{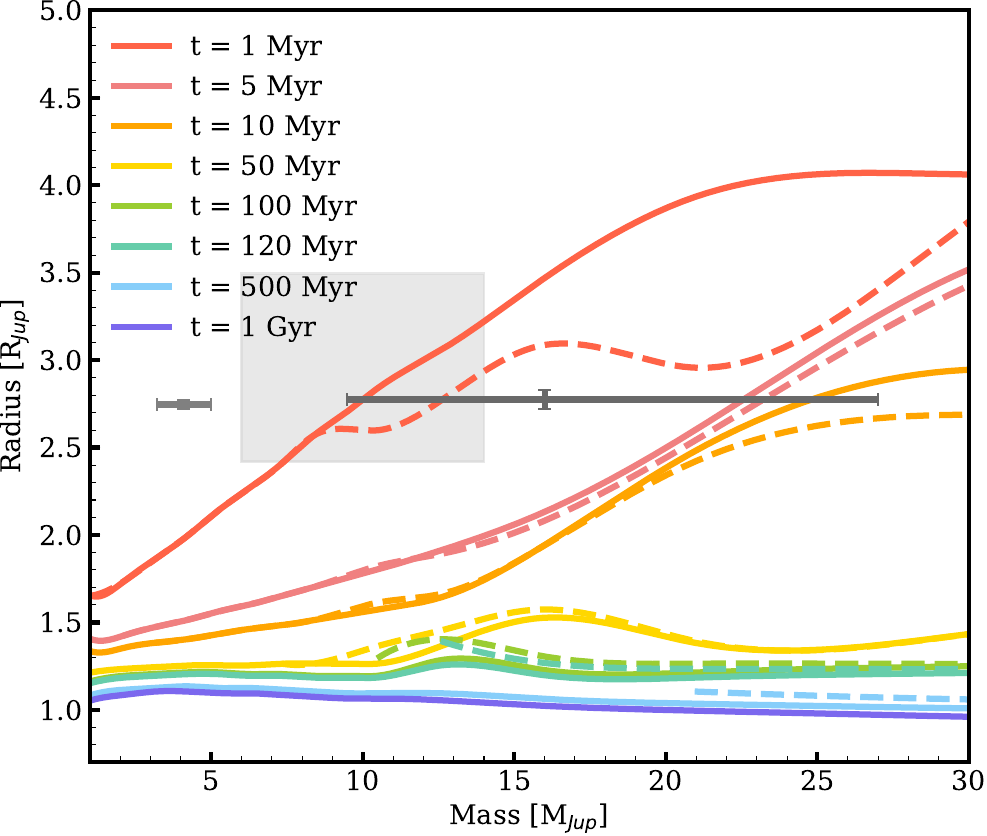}
\caption{Radius as a function of mass at multiple epochs from multiple evolution model grids: the COND model grid (solid lines) and the DUSTY model grid from \cite{allard_2001a} (dashed lines). The grey box represents the photometry-derived mass and radius from previous studies \citep{currie_2014,daemgen_2017}. The grey point on the left is calculated using the radius and log($g$) values from our joint retrieval. The dark grey point on the right is calculated using the radius and log($g$) values from our low-resolution only retrieval. 
\label{fig:evolution}}
\end{figure} 

\section{Discussion} \label{sec:Discussion}

\subsection{Properties of the ROXs 42B System}

We retrieve a C/O ratio of $0.50\pm0.05$ for ROXs 42B b from our high-resolution fits, in good agreement with the measured C/O ratio of $0.54\pm0.01$ for its host binary, ROXs 42B A+B. We find a similarly consistent sub-solar metallicity of $-0.67\pm0.35$ for the planet and $-0.30\pm0.03$ for the host binary, respectively, although we note that our value for the planet is poorly constrained in the high-resolution retrieval. Our retrieved metallicity for the host binary is in good agreement with the small number of published metallicity measurements for other stellar members of $\rho$-Ophuchius complex, which range from [Fe/H]$=-0.3$ to $+0.05$ \citep{spina_gaia_2017}. Other young star forming regions with more extensive metallicity measurements span a similar range of values \citep{spina_gaia_2017}. The fact that the primary in this system is a close binary is also consistent with a metallicity on the low end of this range. Both surveys and simulations of star formation have shown that the frequency of close binaries increases with decreasing stellar metallicity \citep{moe_close_2019,bate_statistical_2019,moe_impact_2021}.

We next consider whether or not our inferred mass and radius for ROXs 42B b are consistent with the predictions of evolutionary models. Our high-resolution retrievals give a log(\textit{g}) value of $3.49\pm0.48$ which is consistent with model predictions for an object of this age and mass, as well as previously published gravity values for ROXs 42B b, which were derived by matching its photometry with spectroscopic model grids sampling effective temperature and log($g$). As discussed earlier, our high-resolution data are continuum normalized and therefore do not constrain the planet radius, which is derived from the overall luminosity. If we take the retrieved log($g$) and radius from our low-resolution fits and use them to calculate a mass we find a value of $M_p$ = $16^{+11}_{-7}$ M$_{J}$ for our free P-T profile retrievals, and $M_p$ = $6.5^{+1.6}_{-0.3}$ M$_{J}$ for our fixed P-T and $A_V$ retrievals. When we include our high-resolution data in the joint retrievals, this value becomes significantly smaller and more precise at $M_p$ = $4.1^{+0.9}_{-0.9}$ M$_{J}$. We compare this mass and radius to predictions from the COND and DUSTY evolutionary model grids in Figure~\ref{fig:evolution}. As a member of the $\rho$-Ophiucus complex, ROXs 42B has an estimated age between 1-5 Myr \citep{kraus_2013}.  We therefore expect the radius of ROXs 42B b to be significantly larger than that of older objects with equivalent masses. We find our low-resolution retrievals give results consistent with previous studies and evolutionary model grids, while our joint retrievals give a mass and radius that imply a planet even younger than 1 Myr. It is possible that this object is younger than 1 Myr, as planet formation can take several Myr, thus the age of the companion may be younger than the isochronal age derived for stellar members of $\rho$-Ophiuchus. However, given the degeneracies and issues discussed above in both these retrievals, and their consequences on the retrieved radii, we cannot be sure either value reflects the true properties of this system.

\subsection{Constraints on Formation and Migration History}

There have been multiple studies of directly imaged companions that have attempted to leverage their atmospheric compositions to constrain their formation and migration histories. \cite{whiteford_retrieval_2023} found a super-solar C/O ratio of $0.97^{+0.09}_{-0.2}$, and a poorly constrained, but broadly consistent with solar, metallicity of [Fe/H] = $-0.04^{+0.95}_{-0.49}$ for 51 Eri b \citep{macintosh_discovery_2015}, but note possible degeneracies with clouds in their models. \cite{brown-sevilla_revisiting_2023} reported a sub-solar C/O = $0.38\pm0.09$ and a metallicity [Fe/H] = $0.26\pm0.30$ for this same planet, which they argue is evidence in favor of formation via core accretion. However, their argument assumes that the host star has a solar C/O ratio.

The planets in the HR 8799 system are considered strong candidates for formation via gravitational instability, as it is difficult to produce four relatively massive planets around a low metallicity star via core accretion unless it hosted an unusually large protoplanetary disk \citep{molliere_2020,ruffio_deep_2021,wang_chemical_2020}. Atmospheric retrievals of HR 8799 e by \cite{molliere_2020} and HR 8799 b,c,d \citep{ruffio_deep_2021} found that all of the planets in this system have C/O ratios consistent with the stellar value \citep{wang_chemical_2020}, in agreement with expectations for formation via disk instability. \cite{ruffio_deep_2021} did not fit for metallicity for HR 8799 b,c,d, but found that solar metallicity templates were consistent with their data. HR 8799 is a $\lambda$ Bo\"{o}tis type star, with a highly sub-solar iron abundance of [Fe/H] = $-0.65^{+0.02}_{-0.01}$ \citep{swastik_host_2021}, and C/H and O/H ratios consistent with solar values \citep{wang_chemical_2020}. Retrievals by \cite{molliere_2020} find a moderate metallicity enhancement of $0.48^{+0.25}_{-0.29}$ for HR 8799 e, the inner planet of this system, which has a similar orbital distance to 51 Eri b at 16 au. There is evidence that the recently discovered AF Lep b \citep{franson_astrometric_2023,de_rosa_direct_2023,mesa_af_2023}, orbiting at around only 9 au, has a substantial metal enrichment ([Fe/H] $>$ 1) relative to its host star ([Fe/H] = $-0.27\pm0.31$), indicating substantial planetismal accretion, consistent with the observed debris disk around the star \citep{zhang_elemental_2023}.  
\begin{figure}[t!]
\epsscale{1.2}
\plotone{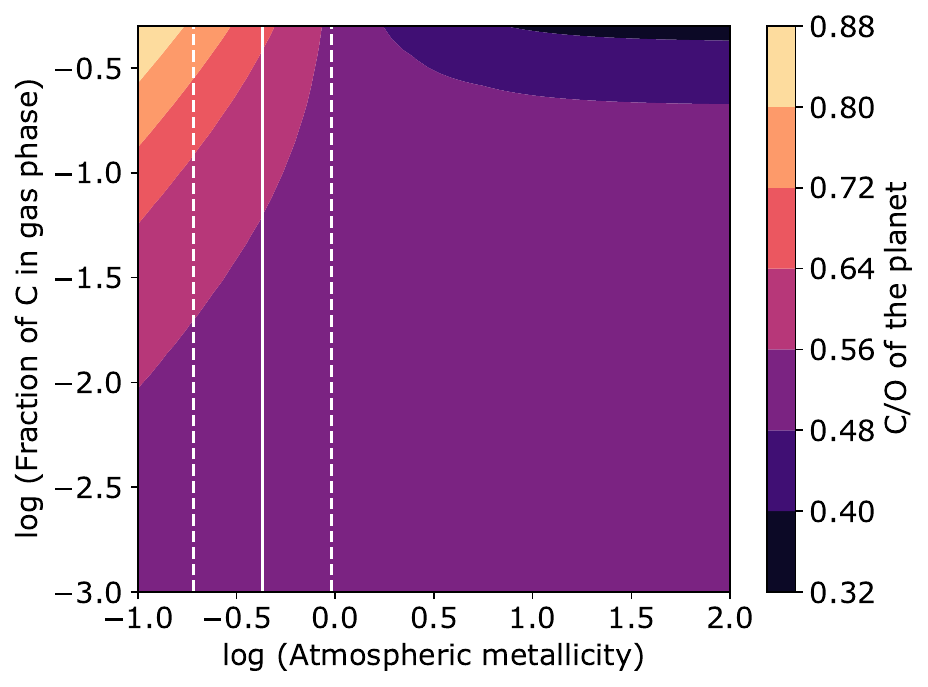}
\caption{Predicted C/O ratio of ROXs 42B b as a function of final atmospheric metallicity and the carbon abundance of the gas accreted by the planet, assuming formation outside the CO snowline. Our derived metallicity constraints from our high-resolution fits are shown as white lines.
\label{fig:solid_accretion}}
\end{figure} 

Our results indicate that ROXs 42B b's C/O ratio and metallicity are also consistent with the stellar value, suggesting that it may share the same formation mechanism as the outer planets of the HR 8799 system. ROXs 42B b orbits at a distance much greater than any of the HR 8799 planets, which are all within 70 au. The CO ice line for an A type star like HR 8799 could lie anywhere between $\sim$50-hundreds of au depending on the disk properties \citep{oberg_effects_2011,qi_chemical_2015,ruffio_deep_2021}, so it is possible the planets in this system could have formed beyond it and then migrated inward afterward.

While it is often assumed that stellar C/O ratios are more likely when planets form via gravitational instability, there is one important caveat to this picture. At large separations where virtually all of the carbon- and oxygen-bearing species have condensed out into solids, the solid component of the gas will have a roughly solar C/O ratio. The gas in this region will be metal-poor, with a C/O ratio of $\sim$1 due to a small amount of residual CO in the gas phase. As a result, planets formed in this region should have stellar C/O ratios as long as they accrete any amount of solids, whether they form from core accretion or disk instability \citep{turrini_tracing_2021,chachan_breaking_2023}. In Fig.~\ref{fig:solid_accretion} we show the expected atmospheric C/O ratio for ROXs 42B b as a function of final atmospheric metallicity, and the carbon abundance of the gas accreted, assuming formation beyond the C/O snowline. For our retrieved atmospheric metallicity constraints, we find that non-stellar C/O ratios are possible, but only for much larger than expected C abundances in the gas phase.

 Our observations may also be consistent with a scenario in which the planet accreted most of its mass at smaller orbital separations before migrating outward and accreting its uppermost atmospheric layers beyond the CO ice line. If the planet's atmospheric envelope is not well mixed and instead has a layered structure similar to that of Jupiter \citep{debras_new_2019,stevenson_jupiters_2020}, the observed C/O ratio would only reflect the material accreted towards the end of the planet formation process. It is possible that dynamical interactions with the primary binary during and after formation (e.g. Kozai-Lidov cycles) could have caused ROXs 42B b to migrate outward, akin to the dynamical evolution of a triple star system. In these three body systems, the orbit of the third body is unstable for orbital separations of less than $\sim$3 times the orbital separation of the primary binary \citep{tokovinin_formation_2020}. Using the present day projected binary separation, we find that the orbit of ROXs 42B b would have become unstable if it was located within $\sim$ 30 au of the primary binary; this is well outside the predicted location of the CO ice line. While it is possible that ROXs 42B b formed in-situ at its currently observed location of 140 au \citep{bryan_2018}, the low expected disk surface densities at these large separations make it difficult to form such a massive planet through any formation channel. If it instead formed closer in (but still beyond 30 au) and then migrated outward, this would also be consistent with our observed C/O ratio.

The atmospheric metallicities of these objects provide us with complementary constraints on their formation and migration histories. Disk instability should produce planets with stellar compositions, although the presence of pressure bumps and spiral structures \citep[e.g.][]{boley_heavy-element_2011} can result in localized enhancements or reductions of metals compared to the surrounding nebula. Additionally, disk instabilities are most likely to occur early on, when the disk is more massive.  If the protoplanet remains embedded in the disk after its formation, it could accrete additional solids and become enriched in metals. This may be more challenging for relatively massive gas giants like ROXs 42B b, which would need to accrete a significant fraction of the total solid budget of the disk in order to appreciably enhance its atmospheric metallicity ($\sim$ 2 M$_J$ of solids for 10$\times$ metallicity predicted by our low-resolution retrievals). In the solar system, there is an ongoing debate over how to supply Jupiter with enough solids to produce its relatively modest metallicity of 3$\times$ solar \citep[e.g.,][]{shibata_enrichment_2022}. Planets that migrate through the disk are predicted to end up with greater metal enhancements, as both pebble accretion \citep{humphries_changes_2018} and planetesimal accretion \citep{shibata_origin_2020,turrini_tracing_2021} become more efficient when the planet moves radially through the disk. \cite{morbidelli_situ_2023} show that giant planets forming in the inner regions of the disk can incorporate significant quantities of dust into their atmospheres, but this process becomes much less efficient in the outer disk where pressure bumps and gaps cut off the supply of pebbles to the giant planet. 

Our retrieved metallicity for ROXs 42B b is both subsolar and slightly sub-stellar, consistent with expectations for in situ formation via gravitational instability. Low metallicity gas cools more quickly due to the lower gas opacity, and low metallicity disks are therefore more likely to fragment \citep{moe_close_2019}. Given the small ($\sim$10 au) orbital separation of the primary binary \citep{kraus_2013}, it is likely that the host stars formed via disk fragmentation. This would mean that either the original disk was fairly massive or that it had high stellar accretion rates in order to produce two separate fragmentation events that formed massive companions \citep{tokovinin_formation_2020}.  However, gas in the outer disk is also expected to be metal-poor due to the condensation of most carbon- and oxygen-bearing species, and our observations could therefore also be explained by accretion of metal-poor gas onto a solid core located in the outer disk. Because small solids tend to migrate inward over time, there might be relatively little solid mass left in the outer disk by the time the planet was accreting the outermost layers of its envelope. 

It is also possible that rather than forming from within the circumstellar disk, ROXs 42B b could have form via turbulent fragmentation during the initial cloud collapse stage, making it analogous to wide binary stars. In this case, it is more surprising that it would have a composition similar to its host star, as wide binaries have been observed to have larger chemical variations \citep[e.g.][]{hawkins_2020,behmard_planet_2023}, but not inconsistent with this formation mechanism. However, given the very low mass of this companion, this is a less likely scenario as cloud collapse events occur when there is a significant gas reservoir, meaning that wide companions with extreme mass ratios are rare \citep{fisher_turbulent_2004}. 

Lastly, we consider whether our retrieved radius for ROXs 42B b might provide additional constraints on its formation mechanism.  The amount of internal energy at formation is determined by the rate at which the planet forms. `Hot start' models assume that the planets form quickly without sufficient time to cool; this is broadly consistent with a process like disk instability, which is expected to occur on relatively short timescales \cite{burrows_beyond_2003,marley_luminosity_2007}. On the other hand, `cold start' models assume a relatively slow formation process, more analogous to core accretion, though core accretion planets could potentially fall in the hot start regime, given that types of models depend on poorly constrained accretion processes. Both the COND and DUSTY models shown in Fig.~\ref{fig:evolution} fall in the hot start regime. We compare our measured mass and radius for ROXs 42B b to the models presented in \cite{spiegel_spectral_2012} and find that we are unable to distinguish between these scenarios with the current large uncertainties in the planet mass and system age. For masses on the high end of our allowed range we can still match the planet's measured radius with cold start models, while masses on the lower end of our range are most consistent with hot start models.

\section{Conclusions}

In this paper we present the results of atmospheric retrievals on the young directly imaged planet ROXs 42B b using low- and high-resolution spectroscopy. We find that our low-resolution spectra are sensitive to the presence of clouds, which are degenerate with the retrieved temperature structure, composition, and radius. We see no sign of this degeneracy in our high-resolution retrievals, which return consistent constraints regardless of our choice of cloud model. We retrieve a C/O ratio of $0.504\pm0.048$ and a subsolar metallicity of $-0.67\pm0.35$ for ROXs 42B b using our high-resolution data. We additionally constrain the projected spin rate $v$sin$i$ to be $10.52\pm0.92$ km s$^{-1}$, which is consistent with previous measurements. We conclude that high-resolution retrievals offer a promising avenue for obtaining robust abundance constraints for cloudy objects like ROXs 42B b. Future high-resolution studies of this planet would benefit from broader wavelength coverage, which should provide tighter constraints on cloud properties.

We also derive abundances for the unresolved binary host, ROXs 42B, using low-resolution SNIFS optical spectroscopy and the SPHINX M dwarf model grid from \cite{iyer_sphinx_2023}. We retrieve a C/O ratio of $0.54\pm0.01$ and metallicity of $-0.30\pm0.03$ for the binary, which are consistent with our retrieved planetary values within 1$\sigma$. This result is consistent with expectations for both disk instability and core accretion  models given ROXs 42B b's wide orbital separation, as it likely formed beyond the CO snow line where all major carbon- and oxygen-carrying species have condensed out of the gas phase. However, the low retrieved metallicity of both the host star and planet, as well as the retrieved C/O ratio, is most consistent with formation in the outer disk via gravitational instability. Measurements of the orbital eccentricity or obliquity of ROXs 42 B b would provide another dimension that could help to better constrain its origins, as high eccentricities and misaligned obliquities are predicted to be common for planets that form via gravitational instability \citep{bowler_population-level_2020,bryan_obliquity_2021}. While both an eccentricity \citep{bowler_population-level_2020} and variability measurement \citep{bowler_2020} for this system exist, they are both only upper limits. Mapping out the orbit of the primary binary would additionally provide extra information on the geometry of this system.

If we wish to obtain better constraints on the cloud properties of ROXs 42B b, mid infrared coverage with JWST/MIRI (5-16 $\mu$m) could provide access to features diagnostic of individual clouds species \citep{wakeford_2015}, including silicate features between 8-10 $\mu$m.  A direct measurement of this silicate feature could help place tighter constraints on the location and scattering properties of cloud particles in ROXs 42B b's atmosphere, potentially mitigating the cloud-temperature degeneracy in the current near-infrared low-resolution data. Similar features were observed in 2M2224-0158 with Spitzer, which \cite{burningham_2021} were able to reproduce in their models using a combination of silicate and quartz clouds. Absorption features from silicate cloud have also be observed in the companion VHS 1256–1257 b with JWST \citep{miles_jwst_2023}.  These observations demonstrate the power of mid-infrared observations in constraining clouds in young isolated objects.

\section*{Acknowledgements}

This research was carried out at the Jet Propulsion Laboratory and the California Institute of Technology under a contract with the National Aeronautics and Space Administration and funded through the President’s and Director’s Research $\&$ Development Fund Program. The computations presented here were conducted in the Resnick High Performance Center, a facility supported by Resnick Sustainability Institute at the California Institute of Technology. Some of the data presented herein were obtained at the W. M. Keck Observatory, which is operated as a scientific partnership among the California Institute of Technology, the University of California and the National Aeronautics and Space Administration. The Observatory was made possible by the generous financial support of the W. M. Keck Foundation. The authors wish to recognize and acknowledge the very significant cultural role and reverence that the summit of Maunakea has always had within the indigenous Hawaiian community.  We are most fortunate to have the opportunity to conduct observations from this mountain.

\software{\texttt{numpy} \citep{harris2020array}, \texttt{scipy} \citep{2020SciPy-NMeth}, \texttt{matplotlib} \citep{Hunter:2007}, \texttt{astropy} \citep{astropy:2013,astropy:2018,astropy:2022}, \texttt{dynesty} \citep{sergey_koposov_2023_7995596}, \texttt{pymultinest}, \texttt{pycuba}, \citep{buchner_multi}, \citep{2020SciPy-NMeth}, \texttt{petitRADTRANS} \citep{molliere_petitradtrans_2019}}

\bibliography{ref}{}
\bibliographystyle{aasjournal}

\end{document}